\documentclass[thmsa]{article}
\usepackage{sw20lart}

\input{tcilatex}
\begin{document}

\author{S. Manoff \thanks{%
Work supported by the National Science Foundation of Bulgaria under Grant
No. F-642}}
\title{\textbf{LAGRANGIAN FORMALISM FOR TENSOR FIELDS}}
\date{In \textbf{Topics in complex analysis, differential geometry}\\
\textbf{and mathematical physics}\\
St. Dimiev and K. Sekigawa eds.,\\
\textit{World Scientific Publishing Co. }\\
\textit{Singapore, 1997, } pp. 177-218}
\maketitle

\begin{abstract}
The Lagrangian formalism for tensor fields over differentiable manifolds
with different (not only by sign) contravariant and covariant affine
connections and a metric [$(\overline{L}_n,g)$-spaces] is considered. The
functional variation and the Lie variation of a Lagrangian density,
depending on components of tensor fields (with finite rank) and their first
and second covariant derivatives are established. A variation operator is
determined and the corollaries of its commutation relations with the
covariant and the Lie differential operators are found. The canonical
(common) method of Lagrangians with partial derivatives (MLPD) and the
method of Lagrangians with covariant derivatives (MLCD) are outlined. They
differ from each other by the commutation relations the variation operator
has to obey with the covariant and the Lie differential operator. The
canonical and covariant Euler-Lagrange equations are found as well as their
corresponding $(\overline{L}_n,g)$-spaces. The energy-momentum tensors are
considered on the basis of the Lie variation and the covariant Noether
identities.
\end{abstract}

\section{Introduction}

The recent development of the models for description of the gravitational
interaction (Hecht and Hehl 1991), (Hehl et al 1995) give rise to a tendency
in using more general differentiable manifolds than the (pseudo) Riemannian
spaces (Eddington 1925), (Schr\"odinger 1950).

The fact that affine connection, which in a point or over a curve in a
Riemannian space can vanish [leading to the principle of equivalence in the
Einstein theory of gravitation (ETG)], can also vanish under a special
choice of the basic system in a space with affine connection and metric (von
der Heyde 1975), (Iliev 1992), (Hartley 1995) shows that the equivalence
principle in the ETG is only a corollary of the mathematical apparatus used
in this theory. Therefore, \textit{every differentiable manifold with affine
connection and metric can be used as a model for space-time in which the
equivalence principle holds}. But, if the manifold has two different (not
only by sign) connections for tangent and co-tangent vector fields, the
situation changes and is worth being investigated. The main characteristics
of differentiable manifolds with different (not only by sign) contravariant
and covariant affine connections and a metric [$(\overline{L}_n,g)$-spaces]
have been outlined in (Manoff 1995) [or in (Manoff 1996)].

There are countless numbers of papers concerning the applications of the
Lagrangian formalism in classical field theories over (pseudo) Euclidean and
(pseudo) Riemannian spaces. Since the crucial works of Noether about the
role of the symmetries of a Lagrangian density for finding out local
conserved quantities the most authors consider a Lagrangian formalism using
invariant with respect to the co-ordinate action of a physical system,
constructed by means of a Lagrangian density with certain symmetry
properties [s. for example (Lovelock and Rund 1975)]. On this basis they
have found the Euler-Lagrange equations and the corresponding conserved
quantities. Other authors have used the invariance of a Lagrangian density
with respect to functional variations [s. for example (Schmutzer 1968)].
Both types of investigations have found their applications in classical
field theories over differentiable manifolds with an affine connection and a
metric. But both types of considerations are using the common method of
Lagrangians with partial derivatives.

The task of this paper is to show that a Lagrangian theory of tensor fields
can be considered over $(\overline{L}_n,g)$-spaces by means of the common
(canonical) method of Lagrangians with partial derivatives (MLPD) or by
means of the method of Lagrangians with covariant derivatives (MLCD). Both
methods differ to each other. The second one is an entirely covariant,
related to tensor fields over a differentiable manifold method, leading to
independent of the affine connections (and therefore, to independent of the
transport of the tensor fields, gauge invariant) field theoretical
structures.

A Lagrangian formalism for tensor fields over differentiable manifolds with
affine connections and a metric has three essential structures:

(a) \textit{The Lagrangian density}.

(b) \textit{The Euler-Lagrange equations}.

(c) \textit{The energy-momentum tensors}.

In Sec. 2. a Lagrangian density for tensor fields and its properties are
considered as well as its functional and Lie variations. A variation
operator is introduced related to the functional variation and the
corollaries of its commutation relations with the Lie differential operator
and with the covariant differential operator are summarised.

In Sec. 3. the Euler-Lagrange equations are obtained as a result of the
functional variation of the Lagrangian density.

In Sec. 4. the energy-momentum tensors are found by the use of the Lie
variation of the Lagrangian density. The main objects under the
consideration could be given schematically as follows

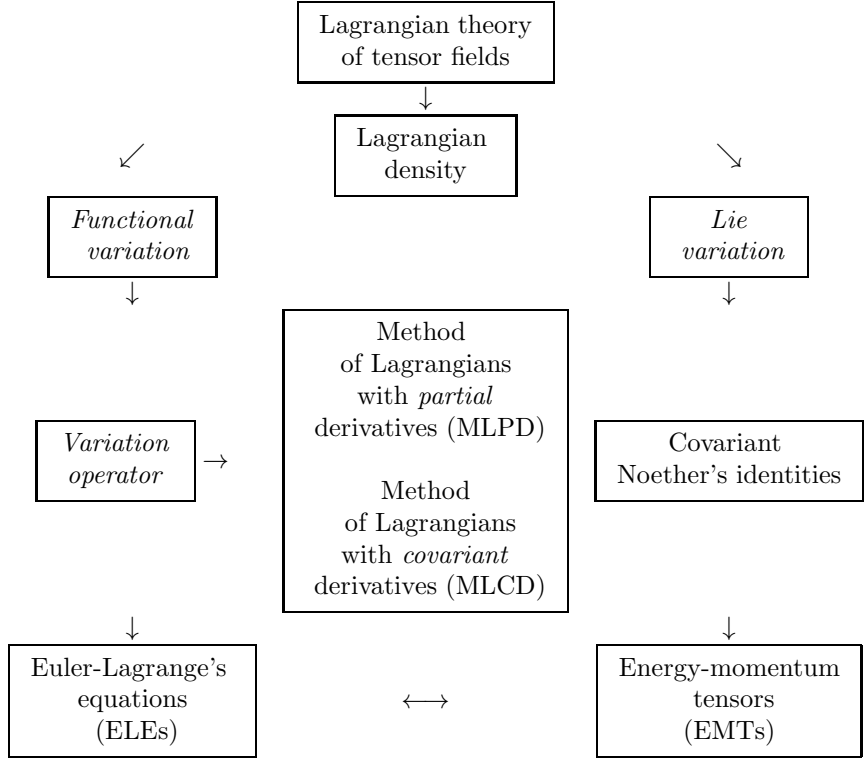
\begin{center}
$
\begin{array}{ccc}
& \fbox{$
\begin{array}{c}
\text{Lagrangian theory} \\ 
\text{of tensor fields}
\end{array}
$} &  \\ 
& \downarrow &  \\ 
\swarrow & \fbox{$
\begin{array}{c}
\text{Lagrangian } \\ 
\text{density}
\end{array}
$} & \searrow \\ 
\fbox{$
\begin{array}{c}
\mathit{Functional} \\ 
\mathit{\,\,\,variation}
\end{array}
$} &  & \fbox{$
\begin{array}{c}
\mathit{Lie} \\ 
\mathit{\ variation}
\end{array}
$} \\ 
\downarrow &  & \downarrow \\ 
\fbox{$
\begin{array}{c}
\mathit{Variation} \\ 
\mathit{\ operator}\text{ }
\end{array}
$}\rightarrow & \fbox{$
\begin{array}{c}
\text{Method } \\ 
\text{of Lagrangians} \\ 
\text{with \textit{partial}} \\ 
\text{ derivatives (MLPD)} \\ 
\\ 
\text{Method} \\ 
\text{ of Lagrangians} \\ 
\text{with \textit{covariant}} \\ 
\text{ derivatives (MLCD)}
\end{array}
$} & \fbox{$
\begin{array}{c}
\text{Covariant } \\ 
\text{Noether's identities}
\end{array}
$} \\ 
\downarrow &  & \downarrow \\ 
\fbox{$
\begin{array}{c}
\text{Euler-Lagrange's } \\ 
\text{equations} \\ 
\text{ (ELEs)}
\end{array}
$} & \longleftrightarrow & \fbox{$
\begin{array}{c}
\text{Energy-momentum} \\ 
\text{ tensors} \\ 
\text{(EMTs)}
\end{array}
$}
\end{array}
$
\end{center}

\section{Lagrangian density $\mathbf{L}$. Functional and Lie variations.
Variation operator}

A Lagrangian density $\mathbf{L}$ is \textit{by definition} a tensor density
(relative tensor field) of rank $0$ with the weight $q=\frac 12$, depending
on tensor fields' components and their first and second covariant
derivatives. In accordance to the two different considerations of a
Lagrangian density $\mathbf{L}$ over a differentiable manifold $M$ ($\dim
M=n $) with affine connections and a metric [$(\overline{L}_n,g)$-space],
one can introduce the following definitions:

(a) \textit{Lagrangian density of type 1.} Tensor density $\mathbf{L}$ with
weight $q=\frac 12$ and rank $0$, considered as depending on components of
tensor fields (with finite rank) and their first (and second) partial
derivatives with respect to the co-ordinates $x^k$ as well as on components
of affine connections and their partial derivatives, 
\begin{equation}  \label{l.1}
\mathbf{L}=\sqrt{-d_g}.L(g_{ij},\,g_{ij,k},\,g_{ij,k,l},\,V^A\,_B,\,V^A%
\,_{B,i},\,V^A\,_{B,i,j})\text{ ,}
\end{equation}

\noindent where $L(x^k)=L^{\prime }(x^{k^{\prime }})$ is a Lagrangian
invariant, $g_{ij}$ are the components of the covariant metric tensor field $%
g=g_{ij}.dx^i.dx^j=g_{\alpha \beta }.e^\alpha .e^\beta $, $dx^i.dx^j=\frac
12(dx^i\otimes dx^j+dx^j\otimes dx^i)$, $g_{ij}=g_{ji}$, $V^A\,_B$ are
components of tensor fields $V$ or components of an affine connection $%
\Gamma $ (with components $\Gamma _{jk}^i$), or $P$ (with components $%
P_{jk}^i$), $d_g=\det (g_{ij})<0$, 
\[
V^A\,_{B,i}=\frac{\partial V^A\text{ }_B}{\partial x^i},\,V^A\,_{B,i,j}=%
\frac{\partial ^2V^A\text{ }_B}{\partial x^j\partial x^i}\text{ .} 
\]

(b) \textit{Lagrangian density of type 2. }Tensor density $\mathbf{L}$ with
weight $q=\frac 12$ and rank $0$, considered as depending on components of
tensor fields (with finite rank) and their first (and second) covariant
derivatives with respect to basic vector fields and to the corresponding
affine connections, 
\begin{equation}  \label{1.2}
\mathbf{L}=\sqrt{-d_g}.L(g_{ij},\,g_{ij;k},\,g_{ij;k;l},\,V^A\,_B,\,V^A%
\,_{B;i},\,V^A\,_{B;i;j})\text{ ,}
\end{equation}

\noindent where $L(x^k)=L^{\prime }(x^{k^{\prime }})$ is a Lagrangian
invariant, $g_{ij}$ are the components of the covariant metric tensor field $%
g$, $V^A\,_B$ are components of tensor fields $V=V^A\,_B.e_A\otimes
e^B=V^A\,_B.\partial _A\otimes dx^B$ with finite rank, $A=i_1...i_k,%
\,B=j_1...j_l,\,k,l\in \mathbf{N}$. $V^A\,_{B;i}$ is the covariant
derivative of the components $V^A\,_B$ along a contravariant basic vector
field $e_i$ (or $\partial _i$) (or along a co-ordinate $x^i$).

The Lagrangian density $\mathbf{L}$ is (by definition) a tensor density of
rank $0$ and weight $\frac 12$ with respect to the parameters $x^m$. On the
other side, $\mathbf{L}$ can be written in the form 
\begin{equation}  \label{1.9}
\mathbf{L}=\mathbf{L}(y(x),w(x),z(x))=\mathbf{L}(r(x))\text{ ,}
\end{equation}

\noindent where 
\begin{equation}
\begin{array}{c}
y\sim (r^1,...,r^s)\text{ , \thinspace \thinspace }w\sim
(r^{s+1},...,r^{s+p})\text{ ,\thinspace \thinspace \thinspace \thinspace }%
z\sim (r^{s+p+1},...,r^{s+p+q})\text{ ,} \\ 
r\sim (r^1,...,r^s,r^{s+1},...,r^{s+p},r^{s+p+1},...,r^{s+p+q})\text{ ,
\thinspace \thinspace \thinspace }s,p,q\in N\text{ .}
\end{array}
\label{1.10}
\end{equation}

$y$ is the set of the functions $K^A\,_B$ (more precisely the image in $R$
of the set of the functions over $R^n$), $w$ is the set of the functions $%
K^A\,_{Bi}$ ($\sim K^A\,_{B;i}$ or $K^A\,_{B,i}$), $z$ is the set of the
functions $K^A\,_{Bij}$ ($\sim K^A\,_{B;i;j}$ or $K^A\,_{B,i,j}$). Every of
the sets of these finite numbers of functions defines a vector space with
the corresponding finite dimension over $M$ ($\dim M=n$). The whole set $%
Y\sim (y,w,z)$ builds a finite vector space as a direct product of the
single vector spaces $y$, $w$ and $z$%
\begin{equation}  \label{1.11}
\begin{array}{c}
Y=y\times w\times z \text{ ,\thinspace \thinspace \thinspace \thinspace }%
Y=R^j\text{ ,\thinspace \thinspace \thinspace \thinspace \thinspace }y=R^s%
\text{ \thinspace ,\thinspace \thinspace \thinspace \thinspace }w=R^{p=n.s}%
\text{ ,\thinspace \thinspace \thinspace \thinspace }z=R^{q=n^2.s}\text{ ,}
\\ 
Y=R^j=R^s\times R^p\times R^q\text{ , \thinspace \thinspace \thinspace
\thinspace }r\in Y\text{ .}
\end{array}
\end{equation}

\textit{Remark}. In accordance with the number of the tensor fields, $y$, $w$
and $z$ could have some linear subspaces.

In this way, the Lagrangian density $\mathbf{L}$ appears as a real function
over $R^j$%
\begin{equation}  \label{1.12}
\mathbf{L}:U(R^j)\rightarrow V(R)\text{ ,\thinspace \thinspace \thinspace
\thinspace \thinspace \thinspace }U(R^j)\subset R^j\text{ ,\thinspace
\thinspace \thinspace \thinspace \thinspace \thinspace }V(R)\subset \mathbf{R%
}\equiv R\text{ .}
\end{equation}

$\mathbf{L}$ could have all properties of a real valued function of a finite
number of variables over a manifold $M$ ($\dim M=n$) and therefore, it could
be considered as an object of the multidimensional analysis. The vector
spaces $y$, $w$ and $z$ appear as linear subspaces of $Y$. Under these
premises the rules and theorems of the multi-dimensional analysis are also
valid for $\mathbf{L}$.

\subsection{The Lagrangian density $\mathbf{L}$ as a differentiable function
over $Y=R^j$}

Let we now consider a real function $\mathbf{L}:U\rightarrow R$, where $%
U\subset Y$.

\textbf{Definition 1.} (Zoritch 1981, p. 435) The function $\mathbf{L}%
:U\rightarrow R$, defined over the set $U\subset R^j$ (or over a
neighbourhood of $r\in U$) is called differentiable at a point $r\in U$ ($r$
is a limit point for $U$) if 
\begin{equation}  \label{2.1}
\mathbf{L}(r+h)-\mathbf{L}(r)=l(r).h+\alpha (r;h)\text{ ,\thinspace
\thinspace \thinspace \thinspace \thinspace \thinspace \thinspace \thinspace 
}h\in R^j\text{ ,}
\end{equation}

\noindent where $l(r):h\rightarrow R$ is a linear function with respect to $%
h $ and $\alpha (r;h)\rightarrow o(h)$ for $h\rightarrow 0$ [$%
\lim_{h\rightarrow 0}[\alpha (r;h)]=0$], $r+h\in U$.

\textit{Remark}. A limit point for $U$ is a point $r$ for which $\overline{r}%
\rightarrow r$, $\overline{r}\in U$. $\overline{r}$ seeks at $r$ over the
set $U$. The basis in $U$ is a set of neighbourhoods of $r$ in the set $U$.
The elements of the basis are denoted as $\hat U_U(r)=U\cap \hat U(r)$.

The vectors 
\begin{equation}  \label{2.2}
\begin{array}{c}
\triangle r(h)=(r+h)-r=h \text{ ,} \\ 
\triangle \mathbf{L}(r;h)=\mathbf{L}(r+h)-\mathbf{L}(r)\text{ }
\end{array}
\end{equation}

\noindent are called correspondingly \textit{increment of the argument} $r$
and \textit{increment of the function} $\mathbf{L}$. These vectors are
usually denoted by the symbols of the functions of $h$: $\triangle r$ and $%
\triangle \mathbf{L}(r)$%
\begin{equation}
\begin{array}{c}
\triangle r:h\rightarrow \triangle r(h)=h\text{ ,} \\ 
\triangle \mathbf{L}(r):h\rightarrow \triangle \mathbf{L}(r;h)=\mathbf{L}%
(r+h)-\mathbf{L}(r)\text{ , \thinspace \thinspace \thinspace \thinspace }%
h\in Y\text{ .}
\end{array}
\label{2.3}
\end{equation}

The linear function $l(r):R^j\rightarrow R$ is called \textit{differential}
(tangential map, derivative map) of the function $\mathbf{L}:U\rightarrow R$
at the point $r\in U$.

The differential of the function $\mathbf{L}:U\rightarrow R$ is usually
marked by means of the symbol $d\mathbf{L}(r)$.

In accordance with the last symbol, \ref{2.1} can be written in the form 
\begin{equation}
\triangle \mathbf{L}(r;h)=\mathbf{L}(r+h)-\mathbf{L}(r)=d\mathbf{L}%
(r).h+\alpha (r;h)\text{ .}  \label{2.4}
\end{equation}

One could see that the differential is in its nature defined over the
increment $h$ of the considered point $r\in R^j$%
\begin{equation}  \label{2.5}
d\mathbf{L}(r).h\in R\text{ ,\thinspace \thinspace \thinspace \thinspace
\thinspace }d\mathbf{L}(r):h\rightarrow d\mathbf{L}(r).h\text{ ,\thinspace
\thinspace }d\mathbf{L}(r):\widetilde{U}\subset R^j\rightarrow R\text{ .}
\end{equation}

\begin{equation}  \label{2.7}
d\mathbf{L}(r):T_r(R^j=Y)\rightarrow T_{\mathbf{L}(r)}(R)\text{ .}
\end{equation}

In full accordance with the one-dimensional case, the function of several
variables $\mathbf{L}$ is differentiable at a point $r$, if its increment $%
\triangle \mathbf{L}(r;h)$ at this point appears as a linear function of the
increment $h$ of the argument $r$ within an accuracy of a term $\alpha (r;h)$
which is infinitesimally small for $h\rightarrow 0$, compared with the
increment $h$ of the argument $r$.

\subsection{Differential and partial derivatives of the Lagrangian density $%
\mathbf{L}$}

Let we now consider the real function $\mathbf{L}:U\rightarrow R$, defined
over the set $U\subset Y=R^j$ and differentiable at the point $r\in U$. In
the most cases $U$ is a range of $Y$. If $r$ is an internal point of the set 
$U$, then the point $r+h$ will (after a sufficient small increment $h$ of $r$%
) also belong to $U$ and therefore, $r+h\in U$ will lie in the domain of $%
\mathbf{L}:U\rightarrow R$. If we express $r$, $h$ and $l(r).h$ in a
co-ordinate basis as 
\begin{equation}  \label{2.8}
\begin{array}{c}
r\sim (r^j)\sim (y^i,w^k,z^l) \text{ ,\thinspace \thinspace \thinspace
\thinspace \thinspace \thinspace }h\sim (h^j)\sim (h^i,h^k,h^l)\text{ ,} \\ 
l(r).h=\overline{a}_j(r).h^j=a_i(r).h^i+b_k(r).h^k+c_l(r).h^l\text{ ,}
\end{array}
\end{equation}

\noindent then the relation $\mathbf{L}(r+h)-\mathbf{L}(r)=l(r).h+\alpha
(r;h)$ can be written in the form 
\begin{equation}
\begin{array}{c}
\mathbf{L}(y^i+h^i,w^k+h^k,z^l+h^l)-\mathbf{L}(y^i,w^k,z^l)= \\ 
=a_i(r).h^i+b_k(r).h^k+c_l(r).h^l+o(h)\text{ \thinspace \thinspace
\thinspace \thinspace \thinspace \thinspace \thinspace (for }h\rightarrow 0%
\text{).}
\end{array}
\label{2.9}
\end{equation}

$a_i(r)$, $b_k(r)$ and $c_l(r)$ are connected to the point $r$ real numbers.

The norm in $R^j$ can be defined in the usual form (Zoritch 1981, p.432), as 
\begin{equation}  \label{2.10}
\parallel r\parallel \,=\sqrt{\sum_i(y^i)^2+\sum_k(w^k)^2+\sum_l(z^l)^2}%
\text{ .}
\end{equation}

Let we now determine the real numbers $a_i(r)$, $b_k(r)$ and $c_l(r)$. For
finding a solution of this problem we will concentrate our attention not on
an arbitrary increment $h$ of $r$ but on a special increment $h^m\in R^s$
colinear to the vector $e_m$ in$\,R^s$. For $h=h^m$ it is obvious that $%
\parallel h\parallel \,=\,\mid h^m\mid $ and we obtain from \ref{2.9} 
\begin{equation}  \label{2.11}
\begin{array}{c}
\mathbf{L}(y^1,...,y^m+h^m,...,y^s,w^k,z^l)-\mathbf{L}%
(y^1,...,y^m,...,y^s)=a_m(r).h^m+o(h)\,\,\, \\ 
\,\,\text{for }h^m\rightarrow 0\text{ .}
\end{array}
\end{equation}

The last equality means that if all variables of the function $\mathbf{L}%
(y^i,w^k,z^l)$ with an exception of $y^m$ are fixed, then the constructed in
this way function of the variable $y^m$ is differentiable at the point $y^m$%
. It follows from \ref{2.11} that 
\begin{equation}  \label{2.12}
a_m(r)=\lim _{h^m\rightarrow 0}\frac 1{h^m}[\mathbf{L}%
(y^1,...,y^m+h^m,...,y^s,w^k,z^l)-\mathbf{L}(y^1,...,y^m,...,y^s,w^k,z^l)]%
\text{ .}
\end{equation}

\textbf{Definition 2.} The limit \ref{2.12} is called \textit{partial
derivative of the function} $\mathbf{L}(r)$ at the point $r\sim
(y^i,w^k,z^l) $ with respect to the variable $y^m$. This derivative is
marked by the symbol 
\[
\frac{\partial \mathbf{L}}{\partial y^m}(r)\text{ .} 
\]

\textbf{Proposition 1.} (Zoritch 1981, p. 437-438). \textit{If the function }%
$\mathbf{L}:U\rightarrow R$\textit{\ defined over a set }$U\subset Y=R^j$%
\textit{\ is differentiable at an internal point }$r\in U$\textit{, then the
function }$\mathbf{L}$\textit{\ has at this point partial derivatives with
respect to every one of the variables and the differential of the function }$%
\mathbf{L}$\textit{\ is uniquely determined by these partial derivatives in
the form} 
\begin{equation}  \label{2.13}
d\mathbf{L}(r).h=\frac{\partial \mathbf{L}}{\partial r^j}(r).h^j=\frac{%
\partial \mathbf{L}}{\partial y^i}(r).h^i+\frac{\partial \mathbf{L}}{%
\partial w^k}(r).h^k+\frac{\partial \mathbf{L}}{\partial z^l}(r).h^l\text{ .}
\end{equation}

\textit{Example}. For the function $\pi ^i:x\rightarrow x^i$, $x^i\in R$, $%
x\in R^n$, i. e. $\pi ^i$ is a maping of $x=(x^1,...,x^n)$ to the
corresponding co-ordinate $x^i$, one has 
\[
\triangle \pi ^i(x,\overline{h})=(x^i+\overline{h}^i)-x^i=\overline{h}^i%
\text{ ,} 
\]

because of $\triangle f(x,\overline{h})=f(x+\overline{h})-f(x)$, where $%
f=\pi ^i$. The increment $\triangle \pi ^i$ of the function $\pi ^i$ is only
a linear to $\overline{h}$ function: $\pi ^i(\overline{h})=\overline{h}%
^i=\triangle \pi ^i(\overline{h})$. Therefore, $\triangle \pi ^i(x,\overline{%
h})=d\pi ^i(x,\overline{h})=\triangle \pi ^i(\overline{h})$ and $d\pi
^i(x)=d\pi ^i$ is independent of $x\in R^n$. If we write $x^i(x)$ instead of 
$\pi ^i(x)$, then it follows that $dx^i(x)\overline{h}=dx^i.\overline{h}=%
\overline{h}^i$ [$dx^i.\overline{h}=dx^i(h)$].

By means of the last relations and on the grounds of the expression \ref
{2.13} one can write the differential of an arbitrary function in the form
of a linear combination of the differentials of the co-ordinates of its
argument $x\in R^n$%
\begin{equation}  \label{2.14}
\begin{array}{c}
df(x).h=\partial _if(x).h^i=\partial _if(r).dx^i(h) \text{ , }\forall h\in
R^n\text{ ,} \\ 
df(x)=\partial _if(x).dx^i=f_{,i}(x).dx^i\text{ .}
\end{array}
\end{equation}

Therefore, 
\begin{equation}  \label{2.15}
d\mathbf{L}(r)=\frac{\partial \mathbf{L}}{\partial y^i}(r).dy^i+\frac{%
\partial \mathbf{L}}{\partial w^k}(r).dw^k+\frac{\partial \mathbf{L}}{%
\partial z^l}(r).dz^l\text{ , \thinspace \thinspace \thinspace }h^i=dr^i(h)%
\text{ , }h=dr(h)\text{ ,}
\end{equation}
\begin{equation}  \label{2.16}
d\mathbf{L}(r).h=\frac{\partial \mathbf{L}}{\partial r^j}(r).h^j=\frac{%
\partial \mathbf{L}}{\partial r^j}(r).dr^j(h)\text{ , }d\mathbf{L}(r)=\frac{%
\partial \mathbf{L}}{\partial r^j}.dr^j\text{ .}
\end{equation}

\subsection{Functional variation of the Lagrangian density $\mathbf{L}$}

The differential $d\mathbf{L}(r)$ of the function $\mathbf{L}(r)$ maps an
element $h\in Y=R^j$ into an element $d\mathbf{L}(r).h\in R$%
\[
d\mathbf{L}(r):h\rightarrow d\mathbf{L}(r).h\in R\text{ ,\thinspace
\thinspace \thinspace \thinspace \thinspace \thinspace \thinspace }r,h\in R^j%
\text{ .} 
\]

$d\mathbf{L}(r).h$ can be written in the form \ref{2.13} 
\[
d\mathbf{L}(r).h=\frac{\partial \mathbf{L}}{\partial r^j}(r).h^j=\frac{%
\partial \mathbf{L}}{\partial y^i}(r).h^i+\frac{\partial \mathbf{L}}{%
\partial w^k}(r).h^k+\frac{\partial \mathbf{L}}{\partial z^l}(r).h^l\text{ .}
\]

Let $r$ and $r+h$ be elements of one and the same subset $U\subset Y=R^j$: $%
r $, $r+h\in U_1\subset U\subset Y$. $h$ is interpreted as the increment of $%
r$. Let we now assume that $h=\varepsilon .\delta r$ is constructed by means
of the image $\delta r$ of $r$ under a certain linear mapping (operator) $%
\delta :r\rightarrow \delta r$ with $r$, $\delta r\in U$, \thinspace
\thinspace \thinspace \thinspace $r$, $h=\varepsilon .\delta r\in U_1\subset
U$, $\varepsilon \in [0,1]\subset R$. Then, 
\begin{equation}  \label{3.1}
r+h=r+\varepsilon .\delta r\text{ ,\thinspace \thinspace \thinspace
\thinspace \thinspace \thinspace \thinspace \thinspace }r^j+h^j=r^j+%
\varepsilon .\delta r^j\text{ ,}
\end{equation}
\begin{equation}  \label{3.2}
d\mathbf{L}(r).h=\varepsilon .d\mathbf{L}(r).\delta r=\varepsilon .\frac{%
\partial \mathbf{L}}{\partial r^j}(r).\delta r^j=\varepsilon .\frac{\partial 
\mathbf{L}}{\partial y^i}(r).\delta y^i+\varepsilon .\frac{\partial \mathbf{L%
}}{\partial w^k}(r).\delta w^k+\varepsilon .\frac{\partial \mathbf{L}}{%
\partial z^l}(r).\delta z^l\text{ .}
\end{equation}

$d\mathbf{L}(r).\delta r$ is called \textit{functional variation} $\delta 
\mathbf{L}(r)$ of the Lagrangian density $\mathbf{L}(r).$ The operator
(mapping) $\delta $ is called \textit{variation operator}. In this case the
expression \ref{2.4} would have the form 
\begin{equation}  \label{3.3}
\mathbf{L}(r+\varepsilon .\delta r)-\mathbf{L}(r)=\varepsilon .d\mathbf{L}%
(r).\delta r+\alpha (r;\varepsilon .\delta r)\text{ ,\thinspace \thinspace
\thinspace \thinspace \thinspace }\lim _{\varepsilon .\delta r\rightarrow
0}\alpha (r;\varepsilon .\delta r)=0\text{ . }
\end{equation}

The last expression can also be written in the form 
\begin{equation}  \label{3.4}
\mathbf{L}(r+\varepsilon .\delta r)-\mathbf{L}(r)=\varepsilon .\delta 
\mathbf{L}(r)+\alpha (r;\varepsilon .\delta r)\text{ ,}
\end{equation}

\noindent where 
\begin{equation}
\delta \mathbf{L}(r)=d\mathbf{L}(r).\delta r=\lim_{\varepsilon \rightarrow
0}\{\frac 1\varepsilon [\mathbf{L}(r+\varepsilon .\delta r)-\mathbf{L}(r)]\}%
\text{ ,}  \label{3.5}
\end{equation}

\noindent or 
\begin{equation}
\delta \mathbf{L}(r)=d\mathbf{L}(r).\delta r=\frac{\partial \mathbf{L}}{%
\partial r^j}(r).\delta r^j=\frac{\partial \mathbf{L}}{\partial y^i}%
(r).\delta y^i+\frac{\partial \mathbf{L}}{\partial w^k}(r).\delta w^k+\frac{%
\partial \mathbf{L}}{\partial z^l}(r).\delta z^l\text{ .}  \label{3.6}
\end{equation}

$\delta \mathbf{L}(r)$ could be considered as the image of $\mathbf{L}(r)$
under the action of the variation operator $\delta $. This action of $\delta 
$ on $\mathbf{L}(r)$ is determined by the definition of $\delta \mathbf{L}%
(r) $%
\begin{equation}  \label{3.7}
\delta :\mathbf{L}(r)\rightarrow \delta \mathbf{L}(r)=d\mathbf{L}(r).\delta r%
\text{ ,\thinspace \thinspace \thinspace \thinspace \thinspace }r\text{%
,\thinspace }\delta r\in Y=R^j\text{, \thinspace \thinspace \thinspace }d%
\mathbf{L}(r).\delta r\in R\text{ , \thinspace \thinspace \thinspace }\delta
:R\rightarrow R\text{ .}
\end{equation}

Let we now consider the integral (Lovelock and Rund 1975, pp. 188-189) 
\begin{equation}  \label{3.8}
S(\varepsilon )=\dint\limits_{V_n}\mathbf{L}(r(x)+\varepsilon .\delta
r(x)).d^{(n)}x\text{ ,}
\end{equation}

\noindent where $V_n$ is the volume of a limited region of $R^n$; $r=r(x)$, $%
x\sim (x^1,...,x^n)$; $r(x)$, $\delta r(x)\in Y=R^j$. The increment $%
h=\varepsilon .\delta r(x)$ (respectively $\delta r$) are also depending on
the parameters $x^m$. $\mathbf{L}(r(x))$ is now considered as a function of
the parameters $x^m$ and the integral $S(\varepsilon )$ is considered as a
function of the parameter $\varepsilon $ [$\delta r(x)$ are considered to be
arbitrary \textit{finite} elements of $Y$].

The integral $S(\varepsilon )$ can be developed with respect to the
parameter $\varepsilon $ by the use of the representation \ref{3.3} of $%
\mathbf{L}(r(x))$ as a differentiable function of $r(x)$%
\begin{equation}  \label{3.9}
\begin{array}{c}
S(\varepsilon )=\dint\limits_{V_n} \mathbf{L}(r(x)+\varepsilon .\delta
r(x)).d^{(n)}x= \\ 
=\dint\limits_{V_n}\mathbf{L}(r(x)).d^{(n)}x+\varepsilon
.\dint\limits_{V_n}\delta \mathbf{L}(r(x)).d^{(n)}x+\dint\limits_{V_n}\alpha
(r(x);\varepsilon .\delta r(x)).d^{(n)}x\text{ }
\end{array}
\end{equation}

\noindent with $\lim_{\varepsilon \rightarrow 0}\alpha (r(x);\varepsilon
.\delta r(x))=0$. $S(\varepsilon )$ will have the form 
\begin{equation}
S(\varepsilon )=S(0)+\varepsilon .\delta S+S(\alpha (\varepsilon ))\text{ ,}
\label{3.10}
\end{equation}

\noindent where 
\begin{equation}
\begin{array}{c}
S(0)=\dint\limits_{V_n}\mathbf{L}(r(x)).d^{(n)}x\text{ , \thinspace
\thinspace }\delta S=\dint\limits_{V_n}\delta \mathbf{L}(r(x)).d^{(n)}x\text{
,} \\ 
\text{ \thinspace }S(\alpha (\varepsilon ))=\dint\limits_{V_n}\alpha
(r(x);\varepsilon .\delta r(x)).d^{(n)}x\text{ .}
\end{array}
\label{3.11}
\end{equation}

The integral $S(\alpha (\varepsilon ))$ is considered (as the other
integrals in \ref{3.10}) over a limited region $U_0(x)\subset R^n$ with a
finite volume $V_n$. In this region $U_0$ the function $\alpha
(r(x);\varepsilon .\delta r(x))$ is a limited function of $r(x)$ and $\delta
r(x)$ [because of the differentiability of $\mathbf{L}(r(x))$ at every $%
r(x)\in U(U_0)\subset Y=R^j$] with $\lim _{\varepsilon \rightarrow 0}[\alpha
(r(x);\varepsilon .\delta r(x))]=0$. Therefore, 
\begin{equation}  \label{3.12}
\begin{array}{c}
\lim _{\varepsilon \rightarrow 0}\dint\limits_{V_n}\alpha (r(x);\varepsilon
.\delta r(x)).d^{(n)}x=\dint\limits_{V_n}\lim _{\varepsilon \rightarrow
0}\alpha (r(x);\varepsilon .\delta r(x)).d^{(n)}x=0 \text{ ,} \\ 
\dint\limits_{V_n}o(\varepsilon .\delta r(x)).d^{(n)}x\leq o(\varepsilon ).M%
\text{ ,\thinspace \thinspace \thinspace \thinspace \thinspace \thinspace }%
M<\infty \text{ , }\varepsilon \rightarrow 0\text{ .}
\end{array}
\end{equation}

The value of the integral $S(\alpha (\varepsilon ))$ aims at $\varepsilon =0$
faster then the integral $\delta S$%
\begin{equation}  \label{3.13}
\lim _{\varepsilon \rightarrow 0}[\frac 1\varepsilon
.\dint\limits_{V_n}\varepsilon .\delta \mathbf{L}(r(x)).d^{(n)}x]=\dint%
\limits_{V_n}\delta \mathbf{L}(r(x)).d^{(n)}x=\text{\thinspace }\delta S%
\text{ .}
\end{equation}

The function $S(\varepsilon )$ is differentiable at $\varepsilon =0$, where $%
\delta S$ is the differential of $S(\varepsilon )$ at the point $\varepsilon
=0$. $\varepsilon .\delta S$ is uniquely determined because of the relation
(Zoritch 1981, p.187) 
\[
\lim _{\varepsilon \rightarrow 0}\frac{S(\varepsilon )-S(0)}\varepsilon
=\lim _{\varepsilon \rightarrow 0}(\delta S+\frac{S(\alpha (\varepsilon ))}%
\varepsilon )=\delta S\text{ .} 
\]

$\delta S$ appears as the derivative of $S(\varepsilon )$ at $\varepsilon =0$%
. 
\begin{equation}  \label{3.14}
\delta S=\left( \frac{dS(\varepsilon )}{d\varepsilon }\right) _{\mid
\varepsilon =0}=0\text{ }
\end{equation}

\noindent is the necessary condition for the existence of an extremum of $S$
at the point $\varepsilon =0$. Therefore, the condition 
\begin{equation}
\delta S=\dint\limits_{V_n}\delta \mathbf{L}(r(x)).d^{(n)}x=0  \label{3.15}
\end{equation}

\noindent appears as a necessary condition for the existence of an extremum
of the function $S(\varepsilon )$ at the point $\varepsilon =0$, to which
the point $r(x)\in Y=R^j$ corresponds: $r(x)=\lim_{\varepsilon \rightarrow
0}[r(x)+\varepsilon .\delta r(x)]$.

$\delta S=0$ can also be written in the form 
\begin{equation}  \label{3.16}
\delta S=\dint\limits_{V_n}d\mathbf{L}(r(x)).\delta r.d^{(n)}x=0\text{ .}
\end{equation}

$\delta r(x)\in U\subset Y=R^j$ can be considered as an arbitrary chosen
element of the subset $U\in Y$. It could have fixed values on the shell (the
boundary) $(V_n)$ of the volume $V_n$ and especially, it can vanish on $%
(V_n) $ 
\begin{equation}  \label{3.17}
\delta r_{\mid (V_n)}=0\text{ .}
\end{equation}

The problem of finding out the conditions for $\mathbf{L}(r(x))$ appearing
as local necessary conditions for $\delta S=0$ with $\delta r_{\mid (V_n)}=0$
is related to the simple Lagrangian problem for the case of tensor fields
with finite rank. As a solution of this problem the Euler-Lagrange equations
follow for the variables $r$ as functions of the co-ordinates $x^m$ of a
region of the manifold $M$.

\textit{Remark}. For simplicity, we identify every point $x$ of the
differentiable manifold $M$ with its co-ordinates $x\sim
(x^i,\,\,\,\,i=1,...,n)\in R^n$ in a given map, where $x:p=x\in M\rightarrow
x(p=x)=x\in R^n$ ($x\in M\simeq x\in R^n$). All functions over $x\in M$ are
considered as functions over $x\in R^n$.

\textit{Remark}. The introduction of the differential and the partial
derivatives of the function $\mathbf{L}(r)$ is equivalent to the
introduction of the Frechet derivative for finite normed spaces. The
considerations of the function $S(\varepsilon )$ is analogous to the
introduction of the Gateaux derivative in finite normed spaces (Hutson and
Pym 1980), (Ljusternik and Sobolev 1982).

$\delta S$ is denoted as the \textit{variation of the action} $S$, where the 
\textit{action} $S$ is defined as 
\begin{equation}  \label{3.18}
S=S(0)=\dint\limits_{V_n}\mathbf{L}(r(x)).d^{(n)}x\text{ .}
\end{equation}

If we interpret $\mathbf{L}$ as a Lagrangian density depending on field
variables and $r$ as \textit{field variables}, then we obtain the well known
relations of the variational problem for the Lagrangian density $\mathbf{L}$.

(a) The Lagrangian density $\mathbf{L}$ depends on components (in a given
basis) of tensor fields (with finite rank) and their first and second
partial derivatives with respect to the co-ordinates as well as on
components of the affine connections over $M$ and their partial derivatives 
\begin{equation}  \label{3.19}
\mathbf{L}=\mathbf{L}(K^A\,_B\text{,}\,\,\,K^A\,_{B,i}\text{,\thinspace
\thinspace \thinspace \thinspace \thinspace }K^A\,_{B,i,j})\text{ ,}
\end{equation}

\noindent where $K^A\,_B\sim (g_{ij},\,\,\,V^A\,_B,\,\,\,\Gamma
_{jk}^i,\,\,\,P_{jk}^i)$.

$g_{ij}$ are the components of the covariant metric tensor $g$ in a given
basis, $V^A\,_B$ are components of non-metric tensor fields $V$, $\Gamma
_{jk}^i$ are the components of the contravariant affine connection $\Gamma $
and $P_{jk}^i$ are the components of the covariant affine connection $P$.

As a tensor density of rang $0$ and weight $\frac 12$ the Lagrangian density 
$\mathbf{L}$ can be constructed by means of the metric tensor $g$ in the
form $\mathbf{L}=\sqrt{-d_g}.L$, \thinspace $d_g=\det (g_{ij})<0$. $L$ is
the Lagrangian invariant with respect to the co-ordinate transformations
(with respect to the diffeomorphisms of the manifold $M$) 
\[
L^{\prime }(x^{k^{\prime }})=L(x^k)\text{ .} 
\]
Therefore, the Lagrangian density $\mathbf{L}$ will have the form 
\begin{equation}  \label{3.20}
\mathbf{L}=\sqrt{-d_g}.L(K^A\,_B\text{,}\,\,\,K^A\,_{B,i}\text{,\thinspace
\thinspace \thinspace \thinspace \thinspace }K^A\,_{B,i,j})\text{ }
\end{equation}

\noindent and the variation $\delta \mathbf{L}$ will be written as 
\begin{equation}
\delta \mathbf{L}=\frac{\partial \mathbf{L}}{\partial K^A\,_B}.\delta
K^A\,_B+\frac{\partial \mathbf{L}}{\partial K^A\,_{B,i}}.\delta
(K^A\,_{B,i})+\frac{\partial \mathbf{L}}{\partial K^A\,_{B,i,j}}.\delta
(K^A\,_{B,i,j})\,\text{ ,}  \label{3.21}
\end{equation}

\noindent where 
\[
\frac{\partial \mathbf{L}}{\partial K^A\,_{B,i}}=\frac{\partial \mathbf{L}}{%
\partial (K^A\,_{B,i})}\text{,\thinspace \thinspace \thinspace \thinspace
\thinspace }\frac{\partial \mathbf{L}}{\partial K^A\,_{B,i,j}}=\frac{%
\partial \mathbf{L}}{\partial (K^A\,_{B,i,j})}\text{ .} 
\]

The validity of commutation relations of the type 
\begin{equation}  \label{3.22}
\delta (K^A\,_{B,i})=(\delta K^A\,_B)_{,i}\text{ , \thinspace \thinspace
\thinspace \thinspace \thinspace \thinspace \thinspace \thinspace \thinspace
\thinspace }\delta (K^A\,_{B,i,j})=(\delta K^A\;_B)_{,i,j}
\end{equation}

\noindent is a problem connected with the validity of commutation relations
between the variation operator $\delta $ and the partial derivatives. It
requires additional investigation. Usually it is assumed that \ref{3.22} is
a priory fulfilled (Schmutzer 1968) or follows on the basis of the condition 
$(\partial /\partial x^i)(\delta r)=0$ and the differentiability conditions
for the function $\mathbf{L}$.

(b) The Lagrangian density $\mathbf{L}$ depends on components of tensor
fields (with finite rank) and their first and second covariant derivatives.
In an analogous way as in the case (a), $\mathbf{L}$ can be written in the
form 
\begin{equation}  \label{3.23}
\mathbf{L}=\sqrt{-d_g}.L(K^A\,_B\text{, \thinspace \thinspace \thinspace }%
K^A\,_{B;i}\text{, \thinspace \thinspace \thinspace }K^A\,_{B;i;j})\text{ ,}
\end{equation}

\noindent where 
\[
K^A\,_B\sim (g_{ij}\text{,\thinspace \thinspace \thinspace \thinspace }%
V^A\,_B)\text{ .} 
\]

The variation $\delta \mathbf{L}$ will have the form 
\begin{equation}  \label{3.24}
\delta \mathbf{L}=\frac{\partial \mathbf{L}}{\partial K^A\,_B}.\delta
K^A\,_B+\frac{\partial \mathbf{L}}{\partial K^A\,_{B;i}}.\delta
(K^A\,_{B;i})+\frac{\partial \mathbf{L}}{\partial K^A\,_{B;i;j}}.\delta
(K^A\,_{B;i;j})\,\text{ ,}
\end{equation}

\noindent where 
\[
\frac{\partial \mathbf{L}}{\partial K^A\,_{B;i}}=\frac{\partial \mathbf{L}}{%
\partial (K^A\,_{B;i})}\text{ ,\thinspace \thinspace \thinspace \thinspace
\thinspace }\frac{\partial \mathbf{L}}{\partial K^A\,_{B;i;j}}=\frac{%
\partial \mathbf{L}}{\partial (K^A\,_{B;i;j})}\text{ .} 
\]

The validity of commutation relations of the type

\begin{equation}  \label{3.24a}
\delta (K^A\,_{B;i})=(\delta K^A\,_B)_{;i}\text{ , \thinspace \thinspace
\thinspace \thinspace \thinspace \thinspace \thinspace \thinspace \thinspace
\thinspace }\delta (K^A\,_{B;i;j})=(\delta K^A\;_B)_{;i;j}
\end{equation}

\noindent is a problem connected with the commutation relations between the
variation operator $\delta $ and the covariant derivatives. It requires
additional investigation.

The Euler-Lagrange equations can be obtained by means of the functional
variations of a Lagrangian density and of these of its field variables
considered as \textit{dynamic field variables} (in contrast to the
non-varied field variables considered as fixed and \textit{non-dynamic field
variables}).

\subsection{Lie variation of the Lagrangian density $\mathbf{L}$}

The free choice of $h\in Y=R^j$ (under the prerequisite for $r+h\in U$, $%
r\in U$) in the expressions \ref{2.1} and \ref{2.13} allows us to consider
an other type of variation than the functional variation.

Let we suppose that the increment $h$ is constructed by means of the action
of the Lie differential operator $\pounds _\xi $ on $r$, considered as
tensor fields components, $\pounds _\xi :r\rightarrow \pounds _\xi r\in Y$,
as 
\begin{equation}  \label{4.1}
h=\varepsilon .\pounds _\xi r\text{ ,\thinspace \thinspace \thinspace
\thinspace \thinspace \thinspace }\forall \xi \in T(M)\text{ , }\varepsilon
\in [0,1]\subset R\,.
\end{equation}

$\pounds _\xi r$ are the (well defined) Lie derivatives of the components of
the tensor fields and their \textit{covariant} derivatives. In this case the
increment $h$ appears as an infinitesimal dragging-along an arbitrary given
contravariant vector field $\xi $ over $M$. Then 
\begin{equation}  \label{4.2}
r+h=r+\varepsilon .\pounds _\xi r\text{ ,\thinspace \thinspace \thinspace
\thinspace \thinspace \thinspace \thinspace \thinspace \thinspace }%
r^i+h^i=r^i+\varepsilon .\pounds _\xi r^i\text{ ,}
\end{equation}
\[
d\mathbf{L}(r).h=\varepsilon .d\mathbf{L}(r).\pounds _\xi r=\varepsilon .[%
\frac{\partial \mathbf{L}}{\partial r^j}(r).\pounds _\xi r^j]= 
\]
\begin{equation}  \label{4.3}
=\varepsilon .[\frac{\partial \mathbf{L}}{\partial y^i}(r).\pounds _\xi y^i+%
\frac{\partial \mathbf{L}}{\partial w^k}(r).\pounds _\xi w^k+\frac{\partial 
\mathbf{L}}{\partial z^l}(r).\pounds _\xi z^l]\text{ .}
\end{equation}

$d\mathbf{L}(r).\pounds _\xi r$ is called \textit{Lie variation of the
Lagrangian density} $\mathbf{L}$. It follows from \ref{2.1} and \ref{4.2}
that 
\begin{equation}  \label{4.4}
\mathbf{L}(r+\varepsilon .\pounds _\xi r)-\mathbf{L}(r)=\varepsilon .d%
\mathbf{L}(r).\pounds _\xi r+\alpha (r;\varepsilon .\pounds _\xi r)\text{ ,}
\end{equation}

\noindent where 
\[
\lim_{\varepsilon .\pounds _\xi r\rightarrow 0}[\alpha (r;\varepsilon
.\pounds _\xi r)]=0\text{ , or }\lim_{\varepsilon \rightarrow 0}[\alpha
(r;\varepsilon .\pounds _\xi r)]=0\text{, if }\mid \pounds _\xi r\mid
_{U\subset Y}\,\leq M<\infty \text{ .} 
\]

The Lagrangian density $\mathbf{L}$ considered as a function of the
parameter $\varepsilon $ is differentiable at the point $\varepsilon =0$%
\begin{equation}  \label{4.5}
\lim _{\varepsilon \rightarrow 0}\frac 1\varepsilon [\mathbf{L}%
(r+\varepsilon .\pounds _\xi r)-\mathbf{L}(r)]=d\mathbf{L}(r).\pounds _\xi r%
\text{ .}
\end{equation}

To the point $\varepsilon =0\in R$ corresponds the point $r\in Y=R^j$.

Let we now consider the integral 
\begin{equation}  \label{4.6}
\begin{array}{c}
\overline{S}(\varepsilon )=\dint\limits_{V_n}\mathbf{L}(r(x)+\varepsilon
.\pounds _\xi r(x)).d^{(n)}x\text{ ,\thinspace \thinspace \thinspace
\thinspace }r=r(x)\text{ ,\thinspace \thinspace \thinspace }\pounds _\xi
r=\pounds _\xi r(x)\text{ , } \\ 
r,\,\,\,\,\,\pounds _\xi r\in Y\text{ , \thinspace \thinspace }x\in R^n\text{
,}
\end{array}
\end{equation}

\noindent and develop this integral with respect to $\varepsilon $ by the
use of the representation \ref{3.3} of $\mathbf{L}(r(x))$ as a
differentiable function of $r(x)$%
\begin{equation}
\begin{array}{c}
\overline{S}(\varepsilon )=\dint\limits_{V_n}\mathbf{L}(r(x)).d^{(n)}x+%
\varepsilon .\dint\limits_{V_n}d\mathbf{L}(r(x)).\pounds _\xi r(x).d^{(n)}x+
\\ 
+\dint\limits_{V_n}\alpha (r(x);\varepsilon .\pounds _\xi r(x)).d^{(n)}x=%
\overline{S}(0)+\varepsilon .\overline{S}(\xi )+\overline{S}(\alpha
(\varepsilon ))\text{ ,}
\end{array}
\label{4.7}
\end{equation}

\noindent where 
\[
\begin{array}{c}
\overline{S}(0)=\dint\limits_{V_n}\mathbf{L}(r(x)).d^{(n)}x=S(0)\text{
,\thinspace \thinspace \thinspace }\overline{S}(\xi )=\dint\limits_{V_n}d%
\mathbf{L}(r(x)).\pounds _\xi r(x).d^{(n)}x\text{ ,} \\ 
\text{\thinspace \thinspace }\overline{S}(\alpha (\varepsilon
))=\dint\limits_{V_n}\alpha (r(x);\varepsilon .\pounds _\xi r(x)).d^{(n)}x%
\text{ .}
\end{array}
\]

We define now the Lie derivative of the integral $\overline{S}(\varepsilon )$
at the point $\varepsilon =0$ along an arbitrary given contravariant vector
field $\xi \in T(M)$ as 
\begin{equation}  \label{4.9}
\pounds _\xi [\overline{S}(0)]=\lim _{\varepsilon \rightarrow 0}\frac{%
\overline{S}(\varepsilon )-\overline{S}(0)}\varepsilon =\pounds _\xi 
\overline{S}\text{ .}
\end{equation}

Since $\overline{S}(\varepsilon )-\overline{S}(0)=\varepsilon .\overline{S}%
(\xi )+\overline{S}(\alpha (\varepsilon ))$, we obtain from \ref{4.9} 
\begin{equation}  \label{4.10}
\begin{array}{c}
\pounds _\xi [ \overline{S}(0)]=\lim _{\varepsilon \rightarrow 0}\frac{%
\overline{S}(\varepsilon )-\overline{S}(0)}\varepsilon =\lim _{\varepsilon
\rightarrow 0}\frac 1\varepsilon [\varepsilon .\overline{S}(\xi )+\overline{S%
}(\alpha (\varepsilon ))]=\overline{S}(\xi )\text{ ,} \\ 
\text{because of }\lim _{\varepsilon \rightarrow 0}\frac 1\varepsilon
S(\alpha (\varepsilon ))=0\text{ .}
\end{array}
\end{equation}

Therefore, 
\begin{equation}  \label{4.11}
\pounds _\xi [\overline{S}(0)]=\overline{S}(\xi )=\dint\limits_{V_n}d\mathbf{%
L}(r(x)).\pounds _\xi r(x).d^{(n)}x\text{ .}
\end{equation}

On the other side, the Lie differential operator $\pounds _\xi $ (Yano
1957), (Manoff 1996) acting on the integral $\overline{S}(0)$ has to commute
with the integral, i. e. it has to obey the relation $\pounds _\xi \,\circ
\,\int =\int \,\circ \,\pounds _\xi $, and therefore, 
\begin{equation}  \label{4.12}
\begin{array}{c}
\pounds _\xi [ \overline{S}(0)]=\pounds _\xi \dint\limits_{V_n}\mathbf{L}%
(r(x)).d^{(n)}x=\dint\limits_{V_n}\pounds _\xi [\mathbf{L}(r(x)).d^{(n)}x]=
\\ 
=\pounds _\xi [\overline{S}(0)]=\overline{S}(\xi )=\dint\limits_{V_n}d%
\mathbf{L}(r(x)).\pounds _\xi r(x).d^{(n)}x\text{ .}
\end{array}
\end{equation}

Since 
\begin{equation}  \label{4.13}
\dint\limits_{V_n}\pounds _\xi [\mathbf{L}(r(x)).d^{(n)}x]=\dint%
\limits_{V_n}[\overline{\pounds }_\xi \mathbf{L}(r(x))].d^{(n)}x\text{ ,}
\end{equation}

\noindent where $\overline{\pounds }_\xi \mathbf{L}(r(x))$ is the Lie
derivative of the tensor density\textbf{\ }$\mathbf{L}$ and $\overline{%
\pounds }_\xi $ is the form-invariant Lie differential operator acting on
covariant tensor densities as $\mathbf{L}$. From the relations \ref{4.11} $%
\div $ \ref{4.13}, it follows that 
\begin{equation}
\overline{\pounds }_\xi \mathbf{L}(r(x))\equiv d\mathbf{L}(r(x)).\pounds
_\xi r(x)\text{ ,}  \label{4.14}
\end{equation}

\noindent or 
\begin{equation}
\overline{\pounds }_\xi \mathbf{L}(r(x))\equiv \frac{\partial \mathbf{L}}{%
\partial r^j}(r).\pounds _\xi r^j=\frac{\partial \mathbf{L}}{\partial y^i}%
(r).\pounds _\xi y^i+\frac{\partial \mathbf{L}}{\partial w^k}(r).\pounds
_\xi w^k+\frac{\partial \mathbf{L}}{\partial z^l}(r).\pounds _\xi z^l\text{ .%
}  \label{4.15}
\end{equation}

\textit{Remark}. From the expressions \ref{4.9} $\div $ \ref{4.11}, it
follows that 
\begin{equation}  \label{4.16}
\dint\limits_{V_n}d\mathbf{L}(r(x)).\pounds _\xi
r(x).d^{(n)}x=\dint\limits_{V_n}\overline{\pounds }_\xi \mathbf{L}%
(r(x)).d^{(n)}x\text{ , }
\end{equation}

\noindent where $\overline{\pounds }_\xi \mathbf{L}(r(x))=d\mathbf{L}%
(r(x)).\pounds _\xi r(x)$ is defined as the Lie variation of the Lagrangian
density $\mathbf{L}$. From \ref{4.14} and \ref{4.16}, it follows that 
\textit{the Lie variation of the Lagrangian density }$\mathbf{L}$\textit{\
is equal to the Lie derivative of the Lagrangian density }$\mathbf{L}$
considered as a tensor density on which the form-invariant Lie differential
operator $\overline{\pounds }_\xi $ acts.

If $\mathbf{L}\,$and $r$ are expressed in their explicit forms, then we
obtain for $\forall \xi \in T(M)$ the s. c. Lie identity for the Lagrangian
density $\mathbf{L}$%
\begin{equation}  \label{4.17}
\overline{\pounds }_\xi \mathbf{L}\equiv \frac{\partial \mathbf{L}}{\partial
K^A\,_B}.\pounds _\xi K^A\,_B+\frac{\partial \mathbf{L}}{\partial K^A\,_{B;i}%
}.\pounds _\xi (K^A\,_{B;i})+\frac{\partial \mathbf{L}}{\partial
K^A\,_{B;i;j}}.\pounds _\xi (K^A\,_{B;i;j})\text{ , }
\end{equation}

\noindent where 
\[
K^A\,_B\sim (g_{ij}\text{, \thinspace \thinspace \thinspace }V^A\,_B)\text{ .%
} 
\]

The next problem is the explicit representation of the identity \ref{4.17}
by means of the explicit form of the Lie derivatives of the components and
their first and second covariant derivatives of the metric tensor field $g$
and the non-metric tensor fields $V$. As a result we obtain the covariant
Noether identities for the Lagrangian density $\mathbf{L}$ and the
corresponding energy-momentum tensors. This problem has been already solved
for a Lagrangian theory of tensor fields (with finite rank) over
differentiable manifolds with an affine connection and a metric [$(L_n,g)$%
-spaces] (Manoff 1991).

\subsection{Variation operator}

\textbf{Definition 3.} \textit{Variation (variational) operator. }Operator $%
\delta $, acting on the components of tensor fields in a given basis and
preserving the type of these tensor fields. The result of its action is
identified with the result of the action of an operator (denoted with the
same symbol $\delta $) on tensor fields with the following properties:

1. Action on a tensor field $K$%
\[
\delta :K\rightarrow \delta K, 
\]
$\,\,$%
\[
\begin{array}{c}
K,\,\,\,\,\,\,\,\delta K\in \otimes ^k\,_l(M), \\ 
\delta K=\delta K^A\,_B.e_A\otimes e^B=\delta K^C\,_D.\partial _C\otimes
dx^D, \\ 
\,K=K^A\,_B.e_A\otimes e^B=K^C\,_D.\partial _C\otimes dx^D, \\ 
\,\,K^A\,_B\in C^r(M),\,\,\,\,\,\,\,\,\,\,\,\,\,\,\,\,\delta K^A\,_B\in
C^r(M),\,x\in M,
\end{array}
\]

(1.1) For $k=l=0$ this property leads to the action od $\delta $ on an
invariant function $f\in \otimes ^0\,_0(M)$: 
\[
\delta :f\rightarrow \delta f\text{ ,\thinspace \thinspace \thinspace
\thinspace \thinspace \thinspace \thinspace \thinspace \thinspace }%
f,\,\,\,\delta f\in C^r(M)\text{ ,\thinspace \thinspace \thinspace
\thinspace \thinspace \thinspace \thinspace \thinspace \thinspace \thinspace 
}f,\,\,\,\,\delta f\in \otimes ^0\,_0(M)\,\text{.} 
\]

\textit{Remark}. A function $f$ over $M$ could be an invariant function $%
[f\in \otimes ^0\,_0(M)$, \thinspace $f\in C^r(M)$] or a noninvariant
function $f\in C^r(M)$, $f\notin \otimes ^0\,_0(M)$ [changing its form and
value under the change of the map (co-ordinates) in $M$].

2. Action on a function $f$ 
\[
\delta :f\rightarrow \delta f\text{ ,\thinspace \thinspace \thinspace
\thinspace \thinspace \thinspace \thinspace \thinspace \thinspace }%
f,\,\,\,\delta f\in C^r(M)\text{ ,} 
\]

3. Linear operator with respect to tensor fields 
\[
\begin{array}{c}
\delta (\alpha .K_1+\beta .K_2)=\alpha .\delta K_1+\beta .\delta K_2 \text{ ,%
} \\ 
\text{ }\alpha ,\beta \in R\text{ (or }C\text{),\thinspace \thinspace
\thinspace \thinspace \thinspace }K_i\in \otimes ^k\,_l(M)\text{ ,
\thinspace \thinspace \thinspace \thinspace }i=1,2\text{ ,}
\end{array}
\]

4. Differential operator acting on tensor fields and obeying the Leibniz
rule 
\[
\begin{array}{c}
\delta (f.g)=\delta f.g+f.\delta g \text{ , \thinspace \thinspace \thinspace
\thinspace \thinspace }f,g\in C^r(M)\text{ ,\thinspace \thinspace \thinspace
\thinspace \thinspace }f,g\in \otimes ^0\,_0\,(M), \\ 
\delta (Q\otimes S)=\delta Q\otimes S+Q\otimes \delta S\text{ ,\thinspace
\thinspace \thinspace \thinspace \thinspace \thinspace }Q\in \otimes
^k\,_l(M)\text{ ,\thinspace \thinspace \thinspace \thinspace }S\in \otimes ^m%
\text{\thinspace }_r(M)\text{ ,}
\end{array}
\]

5a. Commutation relations (as additional conditions) with the Lie
differential operator 
\[
\begin{array}{c}
\delta \circ \pounds _{\partial _j}=\pounds _{\partial _j}\circ \delta \text{
, \thinspace \thinspace \thinspace \thinspace \thinspace \thinspace
\thinspace }\delta \circ \pounds _{e_\alpha }=\pounds _{e_\alpha }\circ
\delta \text{ ,} \\ 
\delta \circ \pounds _\xi -\pounds _{\delta \xi }=\pounds _\xi \circ \delta 
\text{ ,}
\end{array}
\]

5b. Commutation relations (as additional conditions) with the covariant
differential operator 
\[
\begin{array}{c}
\delta \circ \nabla _{\partial _j}=\nabla _{\partial _j}\circ \delta \text{
, \thinspace \thinspace \thinspace \thinspace \thinspace \thinspace
\thinspace \thinspace }\delta \circ \nabla _{e_\alpha }=\nabla _{e_\alpha
}\circ \delta \text{ ,} \\ 
\delta \circ \nabla _\xi -\nabla _{\delta \xi }=\nabla _\xi \circ \delta 
\text{ ,}
\end{array}
\]

5c. Commutation relations (as additional conditions) with the contraction
operator $S$%
\[
\delta \circ S=S\circ \delta \text{ .} 
\]

From the properties 2. and 4. it follows that $\delta 1=0\,,\,\,1\in \mathbf{%
N}\subset C^r(M)$.

\textit{Proof}: $\delta (1.1)=(\delta 1).1+1.(\delta 1)=2.(\delta 1)=\delta
1:\delta 1=0$.

From the properties 2., 3. and 4. it follows that $\delta \alpha =0$%
,\thinspace $\alpha =const.\in R\,$(or $C$) $\subset C^r(M)$.

\textit{Proof}: $\delta (\alpha .g)=\alpha .\delta g=\delta \alpha .g+\alpha
.\delta g:\delta \alpha =0,\,\forall g\in C^r(M)$.

The commutation relations of the variational operator with the covariant
differential operator, the Lie-differential operator and the contraction
operator $S$ could be represented in the following scheme. The commutation
relations of type A appear as sufficient conditions for the commutation of $%
\delta $ with the partial derivatives of the field variables. The
commutation relation of type B appear as sufficient conditions for the
commutation of $\delta $ with the covariant derivative of the field
variables.

\begin{center}
$
\begin{array}{cc}
\begin{array}{c}
\text{Commutation relations of type A} \\ 
\begin{array}{c}
\delta \circ \pounds _{\partial _j}=\pounds _{\partial _j}\circ \delta \text{
, \thinspace }\delta \circ \pounds _{e_\alpha }=\pounds _{e_\alpha }\circ
\delta \text{ ,} \\ 
\delta \circ \pounds _\xi -\pounds _{\delta \xi }=\pounds _\xi \circ \delta
\end{array}
\end{array}
& 
\begin{array}{c}
\text{Commutation relations of type B} \\ 
\delta \circ \nabla _{\partial _j}=\nabla _{\partial _j}\circ \delta \text{
,\thinspace \thinspace }\delta \circ \nabla _{e_\alpha }=\nabla _{e_\alpha
}\circ \delta \text{ ,} \\ 
\delta \circ \nabla _\xi -\nabla _{\delta \xi }=\nabla _\xi \circ \delta
\end{array}
\\ 
\delta (f_{,j})=(\delta f)_{,j}\,,\,\,\delta (e_\alpha f)=e_\alpha (\delta f)
& \delta (f_{,j})=(\delta f)_{,j} \text{ ,\thinspace \thinspace \thinspace
\thinspace }\delta (e_\alpha f)=e_\alpha (\delta f) \\ 
\delta C_{\alpha \beta }\,^\gamma =0\text{ ,\thinspace \thinspace \thinspace
\thinspace \thinspace \thinspace \thinspace }\delta T_{\alpha \beta }^\gamma
=\delta \Gamma _{\beta \alpha }^\gamma -\delta \Gamma _{\alpha \beta }^\gamma
& \delta T_{\alpha \beta }^\gamma =-\,\,\delta C_{\alpha \beta }\,^\gamma \\ 
\begin{array}{c}
\delta (P_{jk}^i+\Gamma _{\underline{j}k}^{\overline{i}})=0 \\ 
\delta (P_{\beta \gamma }^\alpha +\Gamma _{\underline{\beta }\gamma }^{%
\overline{\alpha }}+C_{\underline{\beta }\gamma }\,^{\overline{\alpha }})=0
\\ 
\delta (P_{Bj}^A+\widetilde{\Gamma }_{Bj}^A)=0
\end{array}
& 
\begin{array}{c}
\delta \Gamma _{jk}^i=0 \text{ ,\thinspace \thinspace \thinspace \thinspace
\thinspace }\delta \Gamma _{\beta \gamma }^\alpha =0 \\ 
\delta P_{jk}^i=0 \text{ ,\thinspace \thinspace \thinspace \thinspace
\thinspace }\delta P_{\beta \gamma }^\alpha =0 \\ 
\delta \Gamma _{Bj}^A=0\text{ ,\thinspace \thinspace \thinspace \thinspace }%
\delta P_{Bj}^A=0
\end{array}
\\ 
& 
\begin{array}{c}
\delta (K^A\,_{B;j})=(\delta K^A\,_B)_{;j} \text{ ,\thinspace } \\ 
\delta (K^A\,_{B/\alpha })=(\delta K^A\,_B)_{/\alpha }
\end{array}
\\ 
\begin{array}{c}
\delta (K^A\,_{B,j})=(\delta K^A\,_B)_{,j} \text{ \thinspace } \\ 
\delta (e_\alpha K^A\,_B)=e_\alpha (\delta K^A\,_B)
\end{array}
& 
\begin{array}{c}
\delta (K^A\,_{B,j})=(\delta K^A\,_B)_{,j} \\ 
\delta (e_\alpha K^A\,_B)=e_\alpha (\delta K^A\,_B)
\end{array}
\\ 
\delta (\Gamma _{jk,l}^i)=(\delta \Gamma _{jk}^i)_{,l} & \delta (\Gamma
_{jk,l}^i)=(\delta \Gamma _{jk}^i)_{,l}=0 \text{ ,\thinspace \thinspace
\thinspace }\delta R^i\,_{jkl}=0 \\ 
\delta (e_\sigma \Gamma _{\beta \gamma }^\alpha )=e_\sigma (\delta \Gamma
_{\beta \gamma }^\alpha ) & 
\begin{array}{c}
\delta (e_\sigma \Gamma _{\beta \gamma }^\alpha )=e_\sigma (\delta \Gamma
_{\beta \gamma }^\alpha )=0 \\ 
\delta R^\delta \,_{\alpha \beta \gamma }=-\Gamma _{\alpha \sigma }^\delta
.\delta C_{\beta \gamma }\,^\sigma
\end{array}
\\ 
\begin{array}{c}
\delta (P_{jk,l}^i)=(\delta P_{jk}^i)_{,l} \\ 
\delta (e_\sigma P_{\beta \gamma }^\alpha )=e_\sigma (\delta P_{\beta \gamma
}^\alpha )
\end{array}
& 
\begin{array}{c}
\begin{array}{c}
\delta (P_{jk,l}^i)=(\delta P_{jk}^i)_{,l}=0 \\ 
\delta P^i\,_{jkl}=0
\end{array}
\\ 
\delta (e_\sigma P_{\beta \gamma }^\alpha )=e_\sigma (\delta P_{\beta \gamma
}^\alpha )=0 \\ 
\delta P^\delta \,_{\alpha \beta \gamma }=-P_{\alpha \sigma }^\delta .\delta
C_{\beta \gamma }\,^\sigma
\end{array}
\end{array}
$

$
\begin{array}{cc}
\text{Commutation relations of type A} & \text{\emph{Commutation relations
of type C}} \\ 
&  \\ 
\delta (\xi ^{\overline{i}}\,_{,}\,_{\underline{j}})=(\delta \xi ^{\overline{%
i}})_{,\underline{j}} &  \\ 
\begin{array}{c}
\delta (f^i\,_k.f_j\,^l)=0 \text{ ,\thinspace \thinspace \thinspace
\thinspace \thinspace }\delta (f^\alpha \,_\sigma .f_\beta \,^\gamma )=0 \\ 
\delta f^i\,_j=\delta f^{\overline{i}}\,_{\underline{j}}\text{ , \thinspace
\thinspace \thinspace }\delta f^\alpha \,_\beta =\delta f^{\overline{\alpha }%
}\,_{\underline{\beta }}
\end{array}
& \delta \circ S=S\circ \delta \\ 
\delta P_{jk}^i=-\delta \Gamma _{\underline{j}k}^{\overline{i}} & \delta
f^i\,_j=0 \text{ ,\thinspace \thinspace \thinspace \thinspace }\delta
f^\alpha \,_\beta =0 \\ 
\delta (\xi ^i\,_{,j})=(\delta \xi ^i)_{,j}\text{ ,\thinspace \thinspace
\thinspace }\delta (e_\beta \xi ^\alpha )=e_\beta (\delta \xi ^\alpha ) & 
\delta (f^i\,_{j,k})=(\delta f^i\,_j)_{,k}=0
\end{array}
$
\end{center}

The method using commutation relations of type $A$ is the method of
Lagrangians with partial derivatives (MLPD). The method using commutation
relations of type $B$ is the method of Lagrangians with covariant
derivatives (MLCD). In this case the affine connections appear as
non-dynamic fields variables and the variation commutes simultaneously with
the partial and the covariant derivatives of the tensor fields components.
The commutation relations of type $C$ could be used when the contraction
tensor field $Sr=f^i\,_j.\partial _i\otimes dx^j=f^\alpha \,_\beta .e_\alpha
\otimes e^\beta $ is considered as a (fixed) non-dynamical tensor field or
when $Sr=Kr=g_j^i.\partial _i\otimes dx^j=g_\beta ^\alpha .e_\alpha \otimes
e^\beta $, i.e. when the contraction tensor field $Sr$ is equal to the
Kronecker tensor field $Kr$. In both cases $\delta \circ S=S\circ \delta $
appears as a sufficient condition for $\delta f^i{}_j=0$.

On the basis of the two types 1. and 2. of the Lagrangian density $\mathbf{L}
$ and the commutation relations $A$ and $B$ two types of methods determining
a Lagrangian formalism can be developed:

(a) Method of Lagrangians with partial derivatives (MLPD).

(b) Method of Lagrangians with covariant derivatives (MLCD).

\subsection{Method of Lagrangians with partial derivatives (MLPD)}

The \textit{method of Lagrangians with partial derivatives (MLCD)} is a
Lagrangian formalism for tensor fields based on:

(a) A Lagrangian density $\mathbf{L}$ of type 1.

(b) The action $S$ of a Lagrangian system described by means of the
Lagrangian density $\mathbf{L}$%
\[
S=\dint\limits_{V_n}\mathbf{L}.d^{(n)}x=\dint\limits_{V_n}L.d\omega \text{ ,}
\]

\noindent where $V_n$ is a volume in the manifold $M$ with $\dim M=n$; $%
d\omega =\sqrt{-d_g}.d^{(n)}x$ is the invariant volume element.

(c) The functional variation $\delta S$ of the action $S$ with the condition 
$(\delta S=0)$ for the existence of an extremum 
\begin{equation}  \label{I-1.2b}
\delta S=\delta \dint\limits_{V_n}\mathbf{L}.d^{(n)}x=\dint\limits_{V_n}%
\delta \mathbf{L}.d^{(n)}x=0\text{ .}
\end{equation}

(d) The functional variation of the Lagrangian density $\mathbf{L}$ in the
form 
\[
\delta \mathbf{L}=\frac{\partial \mathbf{L}}{\partial g_{ij}}.\delta g_{ij}+%
\frac{\partial \mathbf{L}}{\partial g_{ij,k}}.\delta (g_{ij,k})+\frac{%
\partial \mathbf{L}}{\partial g_{ij,k,l}}.\delta (g_{ij,k,l})+ 
\]
\begin{equation}
+\frac{\partial \mathbf{L}}{\partial V^A\,_B}.\delta V^A\,_B+\frac{\partial 
\mathbf{L}}{\partial V^A\,_{B,i}}.\delta (V^A\,_{B,i})+\frac{\partial 
\mathbf{L}}{\partial V^A\,_{B,i,j}}\delta (V^A\,_{B,i,j})\text{ .}
\label{I-1.2c}
\end{equation}

(e) The variational operator $\delta $ obeying the commutation relations
with the Lie differential operator 
\begin{equation}  \label{I-1.3}
\begin{array}{c}
\delta \circ \pounds _\xi =\pounds _\xi \circ \delta +\pounds _{\delta \xi } 
\text{ ,} \\ 
\text{\thinspace \thinspace \thinspace \thinspace }\delta \circ \pounds
_{\partial _j}=\pounds _{\partial _j}\circ \delta \text{ \thinspace
\thinspace (in a co-ordinate basis),} \\ 
\delta \circ \pounds _{e_\alpha }=\pounds _{e_\alpha }\circ \delta \,\,\,\,\,%
\text{(in a non-co-ordinate basis),}
\end{array}
\end{equation}

\noindent leading to commutation of $\delta $ with the partial derivatives: 
\begin{equation}
\begin{array}{c}
\delta (g_{ij,k})=(\delta g_{ij})_{,k}\text{ ,\thinspace \thinspace
\thinspace \thinspace \thinspace }\delta (g_{ij,k,l})=(\delta g_{ij})_{,k,l}%
\text{ ,} \\ 
\delta (V^A\,_{B,i})=(\delta V^A\,_B)_{,i}\text{ , \thinspace \thinspace
\thinspace }\delta (V^A\,_{B,i,j})=(\delta V^A\,_B)_{,i,j}\text{ .}
\end{array}
\label{I-1.4}
\end{equation}

(f) The Lie variation $\pounds _\xi S$ of the action $S$%
\begin{equation}  \label{I-1.4a}
\pounds _\xi S=\pounds _\xi \dint\limits_{V_n}\mathbf{L}.d^{(n)}x=\dint%
\limits_{V_n}\overline{\pounds }_\xi \mathbf{L}.d^{(n)}x\text{ \thinspace .}
\end{equation}

(g) The Lie variation of the Lagrangian density $\mathbf{L}$ in the form 
\[
\overline{\pounds }_\xi \mathbf{L}\equiv \frac{\partial \mathbf{L}}{\partial
g_{ij}}.\pounds _\xi g_{ij}+\frac{\partial \mathbf{L}}{\partial g_{ij,k}}%
.\pounds _\xi (g_{ij,k})+\frac{\partial \mathbf{L}}{\partial g_{ij,k,l}}%
.\pounds _\xi (g_{ij,k,l})+ 
\]
\begin{equation}  \label{I-1.5}
+\frac{\partial \mathbf{L}}{\partial V^A\,_B}.\pounds _\xi V^A\,_B+\frac{%
\partial \mathbf{L}}{\partial V^A\,_{B,i}}.\pounds _\xi (V^A\,_{B,i})+\frac{%
\partial \mathbf{L}}{\partial V^A\,_{B,i,j}}.\pounds _\xi (V^A\,_{B,i,j})%
\text{ .}
\end{equation}

\textit{Remark}. In the last expression, the Lie derivatives of the partial
derivatives of the components of tensor fields have to be determined in an
appropriate manner (Schmutzer 1968). There is no covariant definition of the
Lie derivative of partial derivatives of the components of tensor fields.

\subsection{Method of Lagrangians with covariant derivatives (MLCD)}

The \textit{method of Lagrangians with covariant derivatives (MLCD)} is a
Lagrangian formalism for tensor fields based on:

(a) A Lagrangian density $\mathbf{L}$ of type 2.

(b) The action $S$ of a Lagrangian system described by means of the
Lagrangian density $\mathbf{L}$%
\[
S=\dint\limits_{V_n}\mathbf{L}.d^{(n)}x=\dint\limits_{V_n}L.d\omega \text{ ,}
\]

\noindent where $V_n$ is a volume in the manifold $M$ with $\dim M=n$; $%
d\omega =\sqrt{-d_g}.d^{(n)}x$ is the invariant volume element.

(c) The functional variation $\delta S$ of the action $S$ with the condition
for the existence of an extremum 
\begin{equation}  \label{I-1.6}
\delta S=\delta \dint\limits_{V_n}\mathbf{L}.d^{(n)}x=\dint\limits_{V_n}%
\delta \mathbf{L}.d^{(n)}x=0\text{ .}
\end{equation}

(d) The functional variation of the Lagrangian density $\mathbf{L}$ in the
form 
\[
\delta \mathbf{L}=\frac{\partial \mathbf{L}}{\partial g_{ij}}.\delta g_{ij}+%
\frac{\partial \mathbf{L}}{\partial g_{ij;k}}.\delta (g_{ij;k})+\frac{%
\partial \mathbf{L}}{\partial g_{ij;k;l}}.\delta (g_{ij;k;l})+ 
\]
\begin{equation}
+\frac{\partial \mathbf{L}}{\partial V^A\,_B}.\delta V^A\,_B+\frac{\partial 
\mathbf{L}}{\partial V^A\,_{B;i}}.\delta (V^A\,_{B;i})+\frac{\partial 
\mathbf{L}}{\partial V^A\,_{B;i;j}}\delta (V^A\,_{B;i;j})\text{ .}
\label{I-1.7}
\end{equation}

(e) The variational operator $\delta $ obeying the commutation relations
with the covariant differential operator 
\begin{equation}  \label{I-1.8}
\begin{array}{c}
\delta \circ \nabla _\xi =\nabla _\xi \circ \delta +\nabla _{\delta \xi } 
\text{ ,} \\ 
\text{\thinspace \thinspace \thinspace \thinspace }\delta \circ \nabla
_{\partial _j}=\nabla _{\partial _j}\circ \delta \text{ \thinspace
\thinspace (in a co-ordinate basis),} \\ 
\delta \circ \nabla _{e_\alpha }=\nabla _{e_\alpha }\circ \delta \,\,\,\,\,%
\text{(in a non-co-ordinate basis),}
\end{array}
\end{equation}

\noindent leading to commutation of $\delta $ with the covariant
derivatives: 
\begin{equation}
\begin{array}{c}
\delta (g_{ij;k})=(\delta g_{ij})_{;k}\text{ ,\thinspace \thinspace
\thinspace \thinspace \thinspace }\delta (g_{ij;k;l})=(\delta g_{ij})_{;k;l}%
\text{ ,} \\ 
\delta (V^A\,_{B;i})=(\delta V^A\,_B)_{;i}\text{ , \thinspace \thinspace
\thinspace }\delta (V^A\,_{B;i;j})=(\delta V^A\,_B)_{;i;j}\text{ .}
\end{array}
\label{I-1.9}
\end{equation}

(f) The Lie variation $\pounds _\xi S$ of the action $S$%
\begin{equation}  \label{I-1.10}
\pounds _\xi S=\pounds _\xi \dint\limits_{V_n}\mathbf{L}.d^{(n)}x=\dint%
\limits_{V_n}\overline{\pounds }_\xi \mathbf{L}.d^{(n)}x\text{ \thinspace .}
\end{equation}

(g) The Lie variation of the Lagrangian density $\mathbf{L}$ in the form 
\[
\overline{\pounds }_\xi \mathbf{L}\equiv \frac{\partial \mathbf{L}}{\partial
g_{ij}}.\pounds _\xi g_{ij}+\frac{\partial \mathbf{L}}{\partial g_{ij;k}}%
.\pounds _\xi (g_{ij;k})+\frac{\partial \mathbf{L}}{\partial g_{ij;k;l}}%
.\pounds _\xi (g_{ij;k;l})+ 
\]
\begin{equation}  \label{I-1.11}
+\frac{\partial \mathbf{L}}{\partial V^A\,_B}.\pounds _\xi V^A\,_B+\frac{%
\partial \mathbf{L}}{\partial V^A\,_{B;i}}.\pounds _\xi (V^A\,_{B;i})+\frac{%
\partial \mathbf{L}}{\partial V^A\,_{B;i;j}}.\pounds _\xi (V^A\,_{B;i;j})%
\text{ .}
\end{equation}

In the MLCD [because of the commutation relations (e)] \textit{the affine
connections }$\Gamma $\textit{\ and }$P$,\textit{\ and their corresponding
curvature tensors are considered as non-dynamic field variables }$(\delta
\Gamma _{jk}^i=0$, $\delta P_{jk}^i=0$, $\delta R^i{}_{jkl}=0$, $\delta
P^i\,_{jkl}=0)$\textit{.} Therefore, in the MLCD variations of the
components and their covariant derivatives of the covariant metric tensor $g$
and the non-metric tensor fields $V$ are considered for given (fixed) affine
connections and for fixed and determined by them types of transports of the
tensor fields. Of course, the question arises how the affine connections can
be found if not by means of a Lagrangian formalism. The first simple answer
is: the affine connections (or the equations for them as functions of the
co-ordinates in $M$) can be found on the grounds of the MLPD and then the
components of the tensor fields (as functions of the co-ordinates in $M$)
can be determined by means of MLCD. This answer could induce an other
question: why two methods have to be applied when one is enough for finding
out all equations for all dynamic field variables. There are at least two
possible answers to this question: 1. The MLCD ensure the finding out
equations for tensor fields (as dynamic field variables). These equations
are (a) covariant (tensorial) equations and (b) form-invariant (gauge
invariant) equations with respect to the affine connections. The affine
connections could be determined on the grounds of additional conditions and
not exactly by means of a variational principle. 2. The MLPD can ensure the
consideration of the affine connections as dynamic field variables and the
finding out their field equations. It cannot give direct answer for the type
(tensorial or non-tensorial) of the equations obtained for the tensor field
variables and their corresponding energy-momentum quantities. The tensorial
character of quantities and relations concerning the tensor fields variables
has to be proved (which, in general, could be a matter of some difficulty)
(Lovelock and Rund 1975).

In the development of the MLCD problems arise connected with the different
conditions for obtaining Euler-Lagrange's equations.

\section{Euler-Lagrange's equations}

Let a Lagrangian density of the type 2 
\[
\mathbf{L}=\sqrt{-d_g}.L(g_{ij}\text{, }g_{ij;k}\text{, }g_{ij;k;l}\text{, }%
V^A\text{ }_B\text{, }V^A\text{ }_{B;i}\text{, }V^A\text{ }_{B;i;j})\text{ } 
\]

\noindent be given. $g_{ij}$ are the components of the metric tensor field
(metric) $g$, $V^A$ $_B$ are components of non-metric tensor fields $V$ with
finite rank $(A=i_1...i_k$, $B=j_1...j_l)$, $A$, $B$ are multi-indices, $%
d_g=\det (g_{ij})<0$, $_{;k}$ is the covariant derivative (constructed by
means of the affine connection $\Gamma $ and / or $P$) with respect to a
basic vector field $e_k$ (or $\partial _k$), $L$ is the corresponding to $%
\mathbf{L}$ Lagrangian invariant.

The functional variation of $\mathbf{L}$ can be considered under the
following conditions:

1. The metric and the affine connection are in general independent
characteristics of a $(\overline{L}_n,g)$-space.

2. As sufficient condition for the commutation of the functional variation $%
\delta $ with the covariant derivative $_{;k}$ along a basic vector field $%
e_k\,($or $\partial _k)$, i.e. as sufficient condition for 
\begin{equation}  \label{I-2.1}
\delta (g_{ij;k})=(\delta g_{ij})_{;k}\text{ , }\delta (V^A\text{ }%
_{B;k})=(\delta V^A\text{ }_B)_{;k}\text{ ,}
\end{equation}

\noindent appears the condition for the operators $\delta $ and $\nabla
_{\partial _k}$, acting on tensor fields over $M$, 
\begin{equation}
\delta \circ \nabla _{\partial _k}=\nabla _{\partial _k}\circ \delta \text{ ,%
}  \label{I-2.1a}
\end{equation}

\noindent and leading also to the conditions 
\begin{equation}
\delta \Gamma _{jk}^i=0\text{,\thinspace \thinspace \thinspace }\delta
P_{jk}^i=0\text{ ,\thinspace \thinspace \thinspace \thinspace }\delta \circ
e_k=e_k\circ \delta \text{ ,\thinspace }\delta \circ \partial _k=\partial
_k\circ \delta \text{.}  \label{I-2.2}
\end{equation}

Therefore, the contravariant affine connection $\Gamma $ with the components 
$\Gamma _{jk}^i$ and the covariant connection $P$ with the components $%
P_{jk}^i$ appear under these conditions as fixed, \textit{non-dynamic field
variables}.

3. $g_{ij}$ and $V^A$ $_B$ are considered as dynamic field variables. The
functional variation of $\mathbf{L}$ will be found with respect to these
dynamic tensor fields.

4. The functional variation $\delta \omega ^A$ of the covariant basis $%
\omega ^{A\text{ }}$ determining (Lovelock and Rund 1975) the invariant
volume element $d\omega $ 
\[
\begin{array}{c}
d\omega =\frac 1{n!}. \sqrt{-d_g}.\varepsilon _A.\omega ^A=\sqrt{-d_g}%
.d^{(n)}x, \\ 
\omega ^A=dx^{i_1}\wedge ...\wedge dx^{i_n}=e^{i_1}\wedge ...\wedge
e^{i_n},\,\,\,\,\,\,d^{(n)}x=dx^1\wedge ...\wedge dx^n, \\ 
\varepsilon _A=\varepsilon _{i_1...i_n}=\varepsilon _{e_1...e_n}\text{ ,
\thinspace \thinspace \thinspace \thinspace \thinspace \thinspace \thinspace
\thinspace \thinspace \thinspace \thinspace }\delta \varepsilon
_{i_1...i_n}=0\text{ ,}
\end{array}
\]

\noindent is equal to zero 
\begin{equation}
\delta \omega ^A=0\text{ .}  \label{I-2.3}
\end{equation}

\textit{Remark}. The last condition 4. follows from the properties of the
functional operator $\delta $.

The functional variation of $\mathbf{L}$ under the conditions 1$\div $4 can
be considered in two different ways for obtaining the Euler-Lagrange
equations for $g_{ij}$ and $V^A$ $_B$. These two ways are determined by the
different types of terms, separated during the variation of $\mathbf{L}$,
and different conditions for the affine connections $\Gamma $ and $P$ for
obtaining a common divergency term, necessary for the application of the
Stokes theorem.

\subsection{Covariant Euler-Lagrange's equations}

The functional variation of $\mathbf{L}$ with respect to $g_{ij}$ and $%
V^A\,_B$ as components of the dynamic tensor variables $g$ and $V$ can be
written in the forms 
\[
\delta \mathbf{L}=\sqrt{-d_g}.(\overline{\delta }_gL+\delta _vL), 
\]
\begin{equation}  \label{I-2.4}
\delta \mathbf{L}=\sqrt{-d_g}.[(\frac{\delta _gL}{\delta g_{kl}}+\frac
12.L.g^{\overline{k}\overline{l}}).\delta g_{kl}+\frac{\delta _vL}{\delta V^A%
\text{ }_B}.\delta V^A\text{ }_B+j^m\text{ }_{;m}]\text{ ,}
\end{equation}

\noindent where 
\begin{equation}
j^m=\,_gj^m+\,_vj^m\text{ ,\thinspace \thinspace \thinspace \thinspace
\thinspace \thinspace \thinspace }\overline{\delta }_gL=(\frac{\delta _gL}{%
\delta g_{kl}}+\frac 12.L.g^{\overline{k}\overline{l}}).\delta g_{kl}+\,_gj^m%
\text{ }_{;m}=\frac{\overline{\delta }_gL}{\delta g_{kl}}.\delta
g_{kl}+\,_gj^m\text{ }_{;m}\text{ .}  \label{I-2.5}
\end{equation}

The functional variation $(\delta _g\mathbf{L}).d^{(n)}x=\delta _g(L.d\omega
)$ with respect to $g_{ij}$ under the conditions 1$\,\div \,$4 can be
written in the form 
\begin{equation}  \label{I-2.6}
\delta _g(L.d\omega )=(\overline{\delta }_gL).d\omega \text{ .}
\end{equation}

If one has to apply the variational principle to the action $S$%
\begin{equation}  \label{I-2.7}
S=\dint\limits_{V_n}\sqrt{-d_g}.L.d^{(n)}x=\dint\limits_{V_n}\mathbf{L}%
.d^{(n)}x=\dint\limits_{V_n}L.d\omega \text{ ,}
\end{equation}

\noindent with $\delta S=\dint\limits_{V_n}\delta \mathbf{L}.d^{(n)}x=0$ and 
$d^{(n)}x$ as a non-invariant volume element, and $d\omega =\sqrt{-d_g}%
.d^{(n)}x$ as the corresponding invariant volume element, then the covariant
Euler-Lagrange equations for $g_{ij}$, respectively for $V^A$ $_B$, 
\begin{equation}
\frac{\overline{\delta }_gL}{\delta g_{ij}}+P^{ij}=0\Longleftrightarrow 
\frac{\delta _gL}{\delta g_{ij}}=-\frac 12.L.g^{\overline{i}\overline{j}%
}-P^{ij}\text{ , \thinspace \thinspace \thinspace \thinspace \thinspace
\thinspace }\frac{\delta _vL}{\delta V^A\text{ }_B}=-\,\,P_A\,^B\text{ ,}
\label{I-2.8}
\end{equation}

\noindent can be found. This can be done \textit{after} rewriting the scalar
density $\sqrt{-d_g}.j^m\,_{;m}$ as a common divergence $(\sqrt{-d_g}%
.j^m)_{,m}$ of a vector field density $\mathbf{j}^m=\sqrt{-d_g}.j^m$ and
using Stokes' theorem with the boundary conditions 
\begin{equation}
\begin{array}{c}
\delta g_{kl}\mid _{(V_n)}=0,\,\,\delta V^A\,_B\mid _{(V_n)}=0, \\ 
\,\,(\delta g_{kl})_{;i}\mid _{(V_n)}=0\,\simeq \,(\delta g_{kl})_{,i}\mid
_{(V_n)}=0,\text{ (}\simeq \text{because of }\delta g_{kl}\mid _{(V_n)}=0%
\text{ } \\ 
\text{and }\delta P_{jk}^i=0\text{)}, \\ 
\,\,(\delta V^{A\,}\,_B)_{;i}\mid _{(V_n)}=0\simeq (\delta V^A\,_B)_{,i}\mid
_{(V_n)}=0,\,\text{(}\simeq \text{because of }\delta V^A\,_B\mid _{(V_n)}=0%
\text{ } \\ 
\text{and }\delta P_{jk}^i=0\text{)},\text{ }
\end{array}
\label{I-2.16a}
\end{equation}

\noindent for the arbitrary variations $\delta g_{kl}$, $\delta V^A$ $_B$
and their (first) covariant (or partial) derivatives on the shell $(V_n)$ of
the volume $V_n$.

For \textit{every} contravariant vector field $j^i$ the necessary and
sufficient conditions for the existence of the relation 
\begin{equation}  \label{I-2.9}
\sqrt{-d_g}.j^i\text{ }_{;i}=\mathbf{j}^i\text{ }_{,i}=(\sqrt{-d_g}.j^i)_{,i}
\end{equation}

\noindent are the conditions for the components $\Gamma _{jk}^i$ of the
affine connection $\Gamma $%
\begin{equation}
q_i=\Gamma _{ik}^k-(\log \sqrt{-d_g})_{,i}=0\text{ ,\thinspace \thinspace
\thinspace \thinspace \thinspace where }\Gamma _{ik}^k=g_l^k.\Gamma _{ik}^l%
\text{,}  \label{I-2.10}
\end{equation}

\noindent or the equivalent conditions 
\begin{equation}
\begin{array}{c}
q_i=T_{ki}^k-C_{ik}\text{ }^k-\frac 12.g^{\overline{k}\overline{l}%
}.g_{kl;i}+g_k^l.g_{l;i}^k=0\text{ } \\ 
\text{(in a non-co-ordinate basis }\{e_i\}\text{ with }[e_i,e_k]=C_{ik}\text{
}^l.e_l\text{) ,} \\ 
\text{where }T_{jk}^i=\Gamma _{kj}^i-\Gamma _{jk}^i-C_{jk}\,^i\text{, }%
T_{ki}^k=g_l^k.T_{ki}^l\text{ , }C_{ik}\,^k=g_l^k.C_{ik}\,^l\text{ ,} \\ 
g_{l;i}^k=\Gamma _{li}^k+P_{li}^k\text{ , \thinspace \thinspace \thinspace
\thinspace \thinspace \thinspace \thinspace }P_{li}^k-P_{il}^k-C_{li}%
\,^k=U_{li}\,^k\text{ ,}
\end{array}
\label{I-2.11}
\end{equation}
\begin{equation}
\begin{array}{c}
q_i=T_{ki}^k-\frac 12.g^{\overline{k}\overline{l}}.g_{kl;i}+g_k^l.g_{l;i}^k=0%
\text{ } \\ 
\text{(in a co-ordinate basis }\{\partial _i\}\text{) ,} \\ 
\text{where }T_{jk}^i=\Gamma _{kj}^i-\Gamma _{jk}^i\text{ ,\thinspace
\thinspace \thinspace \thinspace \thinspace \thinspace }g_{l;i}^k=\Gamma
_{li}^k+P_{li}^k\text{ , \thinspace \thinspace \thinspace \thinspace
\thinspace \thinspace \thinspace }P_{jk}^i-P_{kj}^i=U_{jk}\,^i\text{ . }
\end{array}
\label{I-2.12}
\end{equation}

The necessary and sufficient conditions follow from the relations 
\begin{equation}  \label{I-2.12a}
\begin{array}{c}
\sqrt{-d_g}.j^i\,_{;i}=(\sqrt{-d_g}j^i)_{,i}+(T_{ki}^k-C_{ik}\,^k-\frac
12.g^{\overline{k}\overline{l}}.g_{kl;i}+g_k^l.g_{l;i}^k).(\sqrt{-d_g}j^i)=
\\ 
= \mathbf{j}^i\,_{,i}+q_i.\mathbf{j}^i\text{ ,\thinspace \thinspace
\thinspace \thinspace }\mathbf{j}^i=\sqrt{-d_g}.j^i\text{ ,\thinspace
\thinspace \thinspace \thinspace \thinspace \thinspace }q_i=T_{ki}^k-C_{ik}%
\,^k-\frac 12.g^{\overline{k}\overline{l}}.g_{kl;i}+g_k^l.g_{l;i}^k \\ 
\text{(in a non-co-ordinate basis),}
\end{array}
\end{equation}
\begin{equation}  \label{I-2.12b}
\begin{array}{c}
\sqrt{-d_g}.j^i\,_{;i}=(\sqrt{-d_g}j^i)_{,i}+(T_{ki}^k-\frac 12.g^{\overline{%
k}\overline{l}}.g_{kl;i}+g_k^l.g_{l;i}^k).(\sqrt{-d_g}j^i)= \\ 
= \mathbf{j}^i\,_{,i}+q_i.\mathbf{j}^i\text{ ,\thinspace \thinspace
\thinspace \thinspace }\mathbf{j}^i=\sqrt{-d_g}.j^i\text{ ,\thinspace
\thinspace \thinspace \thinspace \thinspace \thinspace }q_i=T_{ki}^k-\frac
12.g^{\overline{k}\overline{l}}.g_{kl;i}+g_k^l.g_{l;i}^k \\ 
\text{(in a co-ordinate basis).}
\end{array}
\end{equation}

The conditions in a co-ordinate basis are needed when Stokes theorem is to
be applied (Lovelock and Runnd 1975). (They are identically fulfilled in $%
V_n $-spaces.)

The necessary and sufficient conditions can also be written in the form $%
q_i.j^i=0$. By the use of the \textit{explicit form} of $j^i$ (see below)
and the arbitrariness of $\delta g_{ij}$ and $\delta V^A\,_B$ and their
covariant derivatives $(\delta g_{ij})_{;k}$ and $(\delta V^A\,_B)_{;i}$,
there follow the covariant Euler-Lagrange equations for $g$ and $V$ in the
form: 
\begin{equation}  \label{A.0}
\frac{\delta _gL}{\delta g_{kl}}+\frac 12.L.g^{\overline{k}\overline{l}%
}+P^{kl}=0\text{ ,\thinspace \thinspace \thinspace \thinspace \thinspace }%
\frac{\delta _vL}{\delta V^A\,_B}+P_A\,^B=0\text{ ,}
\end{equation}

\noindent where 
\begin{equation}
\frac{\delta _gL}{\delta g_{kl}}=\frac{\partial L}{\partial g_{kl}}-(\frac{%
\partial L}{\partial g_{kl;m}})_{;m}+(\frac{\partial L}{\partial g_{kl;m;n}}%
)_{;n;m}\text{ ,}  \label{A.01}
\end{equation}
\begin{equation}
P^{kl}=q_i.[\frac{\partial L}{\partial g_{kl;i}}-(\frac{\partial L}{\partial
g_{kl;i;j}}+\frac{\partial L}{\partial g_{kl;j;i}})_{;j}]+(q_i.q_j-q_{i;j}).%
\frac{\partial L}{\partial g_{kl;j;i}}\text{ ,}  \label{A.02}
\end{equation}
\begin{equation}
\frac{\delta _vL}{\delta V^A\text{ }_B}=\frac{\partial L}{\partial V^A\text{ 
}_B}-(\frac{\partial L}{\partial V^A\text{ }_{B;i}})_{;i}+(\frac{\partial L}{%
\partial V^A\text{ }_{B;i;j}})_{;j;i}\text{ ,}  \label{A.03}
\end{equation}
\begin{equation}
P_A\,^B=q_i.[\frac{\partial L}{\partial V^A\,_{B;i}}-(\frac{\partial L}{%
\partial V^A\,_{B;i;j}}+\frac{\partial L}{\partial V^A\,_{B;j;i}}%
)_{;j}]+(q_i.q_j-q_{i;j}).\frac{\partial L}{\partial V^A\,_{B;j;i}}\text{ ,}
\label{A.04}
\end{equation}
\[
\delta _g\mathbf{L=}\sqrt{-d_g}.(\frac{\delta _gL}{\delta g_{kl}}+\frac
12.L.g^{\overline{k}\overline{l}}).\delta g_{kl}+\sqrt{-d_g}._gj^i\,_{;i}= 
\]
\[
=\sqrt{-d_g}.(\frac{\delta _gL}{\delta g_{kl}}+\frac 12.L.g^{\overline{k}%
\overline{l}}+P^{kl}).\delta g_{kl}+(\sqrt{-d_g}._g\overline{j}\,^i)_{,i}%
\text{ ,} 
\]
\[
_g\overline{j}\,^i=\,_gj^i+q_j.\frac{\partial L}{\partial g_{kl;i;j}}.\delta
g_{kl}\text{ ,} 
\]
\[
_gj^m=[\frac{\partial L}{\partial g_{kl;m}}-(\frac{\partial L}{\partial
g_{kl;m;n}}+\frac{\partial L}{\partial g_{kl;n;m}})_{;n}]\delta g_{kl}+(%
\frac{\partial L}{\partial g_{kl;n;m}}.\delta g_{kl})_{;n}\text{ ,} 
\]
\[
\delta _v\mathbf{L}=\sqrt{-d_g}.(\frac{\delta L}{\delta V^A\,_B}.\delta
V^A\,_B+\,_vj^i\,_{;i})= 
\]
\[
=\sqrt{-d_g}.(\frac{\delta L}{\delta V^A\,_B}+P_A\,^B).\delta V^A\,_B+(\sqrt{%
-d_g}.\,_v\overline{j}\,^i)_{,i}\text{ ,} 
\]
\[
_v\overline{j}\,^i=\,_vj^i+q_j.\frac{\partial L}{\partial V^A\,_{B;i;j}}%
.\delta V^A\,_B\text{ ,} 
\]
\[
_vj^m=[\frac{\partial L}{\partial V^A\text{ }_{B;m}}-(\frac{\partial L}{%
\partial V^A\text{ }_{B;m;n}}+\frac{\partial L}{\partial V^A\text{ }_{B;n;m}}%
)_{;n}].\delta V^A\text{ }_B+(\frac{\partial L}{\partial V^A\text{ }_{B;n;m}}%
.\delta V^A\text{ }_B)_{;n}\text{ .} 
\]

\textit{Explicit form of the covariant Euler-Lagrange equations for} $g_{ij}$
\textit{under the conditions} $q_i=0$ \textit{for the affine connections}.

For $q_i=0:P^{kl}=0$ and we have: 
\begin{equation}  \label{A.1}
\delta _g\mathbf{L}=\sqrt{-d_g}.\overline{\delta }_gL=\sqrt{-d_g}(\frac{%
\overline{\delta }_gL}{\delta g_{kl}}.\delta g_{kl}+\,_gj^m\text{ }_{;m})%
\text{ ,}
\end{equation}
\begin{equation}  \label{A.2}
\text{ }\frac{\overline{\delta }_gL}{\delta g_{kl}}=\frac{\delta _gL}{\delta
g_{kl}}+\frac 12.L.g^{\overline{k}\overline{l}}=0\text{ for }\forall \delta
g_{ij}\text{,}
\end{equation}

From the Euler-Lagrange equations \ref{A.2} for $g_{kl}$, it follows that 
\begin{equation}  \label{A.5}
\frac{\delta _gL}{\delta g_{ij}}.g_{ij}=-\frac n2.L\text{ , (}\dim M=n\text{%
).}
\end{equation}

For the special case, when $L=L(g_{ij}$, $V^A$ $_B$, $V^A$ $_{B;i}$, $V^A$ $%
_{B;i;j})$%
\begin{equation}  \label{A.6}
\frac{\delta _gL}{\delta g_{ij}}=\frac{\partial L}{\partial g_{ij}}\text{ ,
\thinspace \thinspace \thinspace \thinspace \thinspace \thinspace \thinspace
\thinspace \thinspace \thinspace \thinspace }\frac{\partial L}{\partial
g_{ij}}.g_{ij}=-\frac n2.L\text{ ,}
\end{equation}

$L$ appears as a homogeneous function of $g_{ij}$ with a degree of
homogeneity $m=-\frac n2$ (Euler's theorem).

\textit{Explicit form of the covariant Euler-Lagrange equations for} $V^A$ $%
_B\, $ \textit{under the conditions }$q_i=0$\textit{\ for the affine
connections.}

For $q_i=0:P_A\,^B=0$ and we have: 
\begin{equation}  \label{A.7}
\delta _v\mathbf{L}=\sqrt{-d_g}.\delta _vL=\sqrt{-d_g}(\frac{\delta _vL}{%
\delta V^A\text{ }_B}.\delta V^A\text{ }_B+\,_vj^m\text{ }_{;m})\text{ ,}
\end{equation}
\begin{equation}  \label{A.8}
\frac{\delta _vL}{\delta V^A\text{ }_B}=\frac{\partial L}{\partial V^A\text{ 
}_B}-(\frac{\partial L}{\partial V^A\text{ }_{B;i}})_{;i}+(\frac{\partial L}{%
\partial V^A\text{ }_{B;i;j}})_{;j;i}=0\text{ .}
\end{equation}

\subsection{Canonical Euler-Lagrange's equations}

The functional variation of the Lagrangian density $\mathbf{L}$ with respect
to $g_{ij}$ and $V^A$ $_B$ can be presented in the form 
\begin{equation}  \label{I-2.13}
\delta \mathbf{L}=\delta _g\mathbf{L}+\delta _v\mathbf{L}=\frac{\delta _g%
\mathbf{L}}{\delta g_{kl}}.\delta g_{kl}+\frac{\delta _v\mathbf{L}}{\delta
V^A\text{ }_B}.\delta V^A\text{ }_B+\mathbf{j}^i\text{ }_{;i}\text{ ,}
\end{equation}

\noindent where 
\begin{equation}
\mathbf{j}^i=\,_g\mathbf{j}^i+\,_v\mathbf{j}^i.  \label{I-2.14}
\end{equation}

The Stokes theorem can be applied for the common divergency $\mathbf{j}%
_{\,\,\,\,,i}^i$ in $(\overline{L}_n,g)$-space $(\dim M=n)$. After
introducing the boundary conditions 
\begin{equation}  \label{I-2.16a}
\begin{array}{c}
\delta g_{kl}\mid _{(V_n)}=0,\,\,\delta V^A\,_B\mid _{(V_n)}=0, \\ 
\,\,(\delta g_{kl})_{;i}\mid _{(V_n)}=0\,\simeq \,(\delta g_{kl})_{,i}\mid
_{(V_n)}=0, \text{ (}\simeq \text{because of }\delta g_{kl}\mid _{(V_n)}=0%
\text{ } \\ 
\text{and }\delta \Gamma _{jk}^i=0\text{),} \\ 
\,\,(\delta V^{A\,}\,_B)_{;i}\mid _{(V_n)}=0\simeq (\delta V^A\,_B)_{,i}\mid
_{(V_n)}=0,\, \text{(}\simeq \text{because of }\delta V^A\,_B\mid _{(V_n)}=0%
\text{ } \\ 
\text{and }\delta \Gamma _{jk}^i=0\text{), }
\end{array}
\end{equation}

\noindent for the arbitrary $\delta g_{kl}$ and $\delta V^A$ $_B$ and their
(first) covariant (or partial) derivatives on the shell $(V_n)$ of the
volume $V_n$, the canonical Euler-Lagrange equations for $g_{kl}$ and $V^A$ $%
_B$ can be found in the form identical to the form of the covariant
Euler-Lagrange equations \ref{A.0}. The result does not depend on the
form-invariant covariant differential operators $\overline{\nabla }_u$ and $%
\widetilde{\nabla }_u$ acting on $\mathbf{L}$ (the definitions of $\overline{%
\nabla }_u$ and $\widetilde{\nabla }_u$ are given below).

If additional conditions are imposed on the affine connections for obtaining
a common divergency term, then the canonical Euler-Lagrangian equations can
be found in the forms respectively:

\begin{equation}  \label{I-2.18}
\frac{\delta _g\mathbf{L}}{\delta g_{kl}}=0\text{ , }\frac{\delta _v\mathbf{L%
}}{\delta V^A\text{ }_B}=0\text{ .}
\end{equation}

\noindent where 
\begin{equation}
\frac{\delta _g\mathbf{L}}{\delta g_{kl}}=\frac{\partial \mathbf{L}}{%
\partial g_{kl}}-(\frac{\partial \mathbf{L}}{\partial g_{kl;m}})_{;m}+(\frac{%
\partial \mathbf{L}}{\partial g_{kl;m;n}})_{;n;m}=0\text{ ,}  \label{I-2.18a}
\end{equation}
\begin{equation}
\frac{\delta _v\mathbf{L}}{\delta V^A\text{ }_B}=\frac{\partial \mathbf{L}}{%
\partial V^A\text{ }_B}-(\frac{\partial \mathbf{L}}{\partial V^A\text{ }%
_{B;k}})_{;k}+(\frac{\partial \mathbf{L}}{\partial V^A\text{ }_{B;k;l}}%
)_{;l;k}=0\text{ .}  \label{I-2.18b}
\end{equation}

The representation of the covariant divergency $\mathbf{j}^i\,_{;i}$ as a
common divergency term $\mathbf{j}^i\,_{,i}$ and additional terms depends on
the form-invariant covariant differential operator by the use of which the
covariant divergency $\mathbf{j}^i\,_{;i}$ has been represented. There exist
two different form-invariant covariant differential operators $\overline{%
\nabla }_u$ and $\widetilde{\nabla }_u$ [$u\in T(M)$] (see the subsection
below about these operators) acting on covariant tensor densities $%
\widetilde{Q}=(d_g)^\omega .Q$ with $\omega =q\in R$ and $Q\in \otimes
^k\,_l(M)$ of the type of $\mathbf{j}^i=\sqrt{-d_g}.j^i$ (here $%
Q=j=j^i.\partial _i=j^\alpha .e_\alpha $, $\omega =q=\frac 12$). The
covariant divergency of $\mathbf{j}^i$ will have the forms

(a) under the action of $\overline{\nabla }_u$:

\begin{equation}  \label{I-2.19}
\begin{array}{c}
\mathbf{j}^i\,_{;i}=\mathbf{j}^i\text{ }_{,i}+(P_{ki}^k+\Gamma _{ik}^k).%
\mathbf{j}^i=\mathbf{j}^i\,_{,i}+(T_{ki}^k-C_{ik}\,^k+g_l^k.g_{k;i}^l).%
\mathbf{j}^i\text{ } \\ 
\text{(in a non-co-ordinate basis),} \\ 
\mathbf{j}^i\,_{;i}=\mathbf{j}^i\text{ }_{,i}+(P_{ki}^k+\Gamma _{ik}^k).%
\mathbf{j}^i=\mathbf{j}^i\,_{,i}+(T_{ki}^k+g_l^k.g_{k;i}^l).\mathbf{j}^i\,
\\ 
\text{(in a co-ordinate basis), }
\end{array}
\end{equation}

(b) under the action of $\widetilde{\nabla }_u$: 
\begin{equation}  \label{I-2.20}
\begin{array}{c}
\mathbf{j}^i\,_{;i}=\mathbf{j}^i\text{ }_{,i}+(P_{ik}^k+\Gamma
_{ik}^k-C_{ik}\,^k).\mathbf{j}^i\text{ \thinspace \thinspace (in a
non-co-ordinate basis),} \\ 
\mathbf{j}^i\,_{;i}=\mathbf{j}^i\text{ }_{,i}+(P_{ik}^k+\Gamma _{ik}^k).%
\mathbf{j}^i\,\,\,\,\,\text{(in a co-ordinate basis). }
\end{array}
\end{equation}

Let we now sketch the properties of the form-invariant covariant
differential operators acting on covariant tensor densities (covariant
relative tensor fields).

\subsection{Tensor densities (relative tensor fields) over $(\overline{L}%
_n,g)$-spaces. Form-invariant covariant differential operators for tensor
densities}

\textbf{1. Tensor densities (Relative tensor fields)}. The notion tensor
density is usually connected with the product of a tensor field with a
determinant to some power of a quadratic matrix. The number of the matrix
elements is equal to the square of the dimension of the manifold. The
elements of the matrix are identified with the components of a tensor field
of second rank, chosen in the most cases to be a covariant metric tensor
field $g$. The necessity of constructing tensor densities is connected with
their transformation properties and with the invariant volume element over a
manifold $M$.

\textbf{2. Contravariant and covariant tensor densities. }For the
construction of a contravariant or covariant tensor densities one can use
the determinant (and its powers) of the components of contravariant or
covariant tensor fields of second rank. It is assumed that these tensor
fields of second rank are symmetric or anti-symmetric tensor fields.

\textbf{Definition 4. }\textit{Contravariant tensor density $\overline{Q}$}
with weight $\omega $ is the product of the tensor field $Q$ with the
determinant $d_{\overline{K}}$ to the power $\omega $ $(\omega =q\in R),$ $%
\overline{Q}=(d_{\overline{K}})^\omega .Q$, where $d_{\overline{K}}=\det (%
\overline{K})=\det (K^{ij})$ (in a co-ordinate basis), $d_{\overline{K}%
}=\det (K^{\alpha \beta })$ (in a non-co-ordinate basis), $d_{\overline{K}%
}\neq 0$,\thinspace \thinspace $Q\in \otimes ^k$ $_l(M)$, $\det (K^{\alpha
\beta })=J^2.\det (K^{\alpha ^{\prime }\beta ^{\prime }})$, $\det (K^{\alpha
^{\prime }\beta ^{\prime }})=J^{-2}.\det (K^{\alpha \beta })$, $J=\det
(A{}_{\alpha ^{\prime }}{}^\alpha )=\det (\partial x^i/\partial x_{i^{\prime
}})$ (in a co-ordinate basis), $J^{-1}=\det (A_\alpha \,^{\alpha ^{\prime
}}) $,$\,\,\,\,\,J.J^{-1}=J^{-1}.J=\mathbf{1}$.

\textbf{Definition 5.} \textit{Covariant tensor density $\widetilde{Q}$}
with weight $\omega $ is the product of the tensor field $Q$ with the
determinant $d_{\underline{K}}$ to the power $\omega $ $(\omega \in R)$, $%
\widetilde{Q}=(d_{\underline{K}})^\omega .Q$, where $d_{\underline{K}}=\det (%
\underline{K})=\det (K_{ij})$ (in a co-ordinate basis), $d_{\underline{K}%
}=\det (K_{\alpha \beta })$ (in a non-co-ordinate basis), $d_{\underline{K}%
}\neq 0$, $\,\,Q\in \otimes ^k$ $_l(M)$, $\det (K_{\alpha \beta
})=J^{-2}.\det (K_{\alpha ^{\prime }\beta ^{\prime }})$, $\det
(K_{ij})=J^{-2}.\det (K_{i^{\prime }j^{\prime }})$.

The transformation properties of tensor densities are determined by the
transformation properties of their constructing quantities: 
\[
\begin{array}{c}
\overline{Q}\,^{\prime }=(d_{\overline{K}}^{\prime })^\omega .Q^{\prime
}=J^{-2\omega }.(d_{\overline{K}})^\omega .Q=J^{-2\omega }.\overline{Q},\,\,
\\ 
\,\,\widetilde{Q}\,^{\prime }=(d_{\underline{K}}^{\prime })^\omega
.Q^{\prime }=J^{2\omega }.(d_{\underline{K}})^\omega .Q=J^{2\omega }.%
\widetilde{Q},\,\,\,\,\,Q^{\prime }=Q.
\end{array}
\]

\textbf{3. Form-invariant covariant differential operators}. From 
\[
\begin{array}{c}
\overline{Q}^{\prime }=J^{-2\omega }.\overline{Q}=J^{-2\omega }.(d_{%
\overline{K}})^\omega .Q\,\,\,\,\,\text{and\thinspace }\widetilde{Q}%
\,^{\prime }=J^{2\omega }.\widetilde{Q}=J^{2\omega }.(d_{\underline{K}%
})^\omega .Q\text{ , \thinspace } \\ 
\text{\thinspace }\,\,\,\,Q=Q_B^A.e_A\otimes e^B\text{ ,}
\end{array}
\]
it follows that $\nabla _u\overline{Q}\,^{\prime }=-2\omega .[u(\log J)].%
\overline{Q}\,^{\prime }+J^{-2\omega }.\nabla _u\overline{Q}$, $\nabla _u%
\widetilde{Q}\,^{\prime }=2\omega .[u(\log J)].\widetilde{Q}\,^{\prime
}+J^{2\omega }.\nabla _u\widetilde{Q}$, and that $\nabla _u\overline{Q}$ and 
$\nabla _u\widetilde{Q}$ do not transform as tensor densities. Therefore,
the result of the action of the covariant differential operator on a tensor
density is not a tensor density with respect to its transformation
properties. New covariant operators have to be constructed mapping a tensor
density in an other tensor density of the same kind. The determination of
such form-invariant covariant differential operators acting on a tensor
density and preserving its type and weight, i. e. mapping a density in a
tensor density of the same type, is possible but not unique.

\textbf{Definition 6.} \textit{Form-invariant covariant differential
operator of type 1. for contravariant tensor densities} $\overline{Q}$.
Covariant differential operator $\overline{\nabla }_u:\overline{\nabla }%
_u=\nabla _u+2\omega .\Gamma _{\beta \gamma }^\beta .u^\gamma $ (in a
non-co-ordinate basis), $\overline{\nabla }_u=\nabla _u+2\omega .\Gamma
_{lk}^l.u^k$ (in a co-ordinate basis)$:\overline{\nabla }_u\overline{Q}=%
\overline{Q}\,^A$ $_{B;k}.u^k.\partial _A\otimes dx^B$, $\overline{Q}=%
\overline{Q}\,^A{}_B.\partial _A\otimes dx^B$, 
\[
\begin{array}{c}
\overline{Q}\,^A{}_{B;k}=(d_{\overline{K}})^\omega .Q^A{}_{B;k}+[(d_{%
\overline{K}})^\omega ]_{;k}.Q^A{}_B= \\ 
=\overline{Q}\,^A{}_{B,k}+\Gamma _{Ck}^A.\overline{Q}\,^C\,_B+P_{Bk}^D.%
\overline{Q}\,^A\,_D+2\omega .\Gamma _{lk}^l.\overline{Q}\,^A\,_B\text{.}
\end{array}
\]
$\overline{Q}\,^A{}_{B;k}$ are called \textit{components of the covariant
derivative of type 1.} of the tensor density $\overline{Q}$ \textit{in a
co-ordinate basis}.

\textbf{Definition 7.} \textit{Form-invariant covariant differential
operator of type 2. for contravariant tensor densities} $\overline{Q}$.
Covariant differential operator $\widetilde{\nabla }_u:\widetilde{\nabla }%
_u=\nabla _u+2\omega .(\Gamma _{\gamma \beta }^\beta +C_{\gamma \beta }$ $%
^\beta ).u^\gamma $ (in a non-co-ordinate basis), $\widetilde{\nabla }%
_u=\nabla _u+2\omega .\Gamma _{kl}^l.u^k$ (in a co-ordinate basis)$:%
\widetilde{\nabla }_u\overline{Q}=\overline{Q}\,^A$ $_{B;k}.u^k.\partial
_A\otimes dx^B$, $\overline{Q}=\overline{Q}\,^A$ $_B.\partial _A\otimes dx^B$%
, 
\[
\begin{array}{c}
\overline{Q}\,^A{}_{B;k}=(d_{\overline{K}})^\omega .Q^A{}_{B;k}+[(d_{%
\overline{K}})^\omega ]_{;k}.Q^A{}_B+2\omega .T_{lk}^l.\overline{Q}\,^A{}_B=
\\ 
=\overline{Q}\,^A\,_{B,k}+\Gamma _{Ck}^A.\overline{Q}\,^C\,_B+P_{Bk}^D.%
\overline{Q}\,^A\,_D+2\omega .\Gamma _{kl}^l.\overline{Q}\,^A\,_B\,\,\text{.}
\end{array}
\]
$\overline{Q}\,_{\,\,\,\,\,\,B;k}^A\,$ are called \textit{components of the
covariant derivative of type 2. of the tensor density} $\overline{Q}$ 
\textit{in a co-ordinate basis}.

\textbf{Definition 8. }\textit{The form-invariant covariant differential
operator of type 1. for covariant tensor densities} $\widetilde{Q}.$
Covariant differential operator $\overline{\nabla }_u:\overline{\nabla }%
_u=\nabla _u+2\omega .P_{ik}^i.u^k$ (in a co-ordinate basis), $\overline{%
\nabla }_u=\nabla _u+2\omega .P_{\beta \gamma }^\beta .u^\gamma $ (in a
non-co-ordinate basis)$:\overline{\nabla }_u\widetilde{Q}=\widetilde{Q}%
\,^A\,_{B;k}.u^k.\partial _A\otimes dx^B$, $\widetilde{Q}=\widetilde{Q}\,^A$ 
$_B.\partial _A\otimes dx^B$, 
\[
\begin{array}{c}
\widetilde{Q}\,_{\,\,\,\,\,\,\,B;k}^A=(d_{\underline{K}})^\omega
.Q_{\,\,\,\,\,\,\,\,B;k}^A+[(d_{\underline{K}})^\omega ]_{;k}.Q^A\,_B= \\ 
=\widetilde{Q}\,_{\,\,\,\,\,\,\,\,B,k}^A+\Gamma _{Ck}^A.\widetilde{Q}%
\,_{\,\,\,\,\,\,B}^C+P_{Bk}^D.\widetilde{Q}\,_{\,\,\,\,\,\,D}^A+2\omega
.P_{lk}^l.\widetilde{Q}\,_{\,\,\,\,\,\,\,B}^A\text{ .}
\end{array}
\]
$\widetilde{Q}\,^A$ $_{B;k}$ are called \textit{components of the covariant
derivative of type 1.} of the tensor density $\widetilde{Q}$ \textit{in a
co-ordinate basis.}

\textbf{Definition 9}. \textit{The form-invariant covariant differential
operator of type 2. for covariant tensor densities of the type of} $%
\widetilde{Q}$ is the covariant differential operator $\widetilde{\nabla }_u:%
\widetilde{\nabla }_u=\nabla _u+2\omega .(P_{\gamma \beta }^\beta -C_{\gamma
\beta }$ $^\beta ).u^\gamma $ (in a non-co-ordinate basis), $\widetilde{%
\nabla }_u=\nabla _u+2\omega .P_{kl}^l.u^k$ (in a co-ordinate basis)$:%
\widetilde{\nabla }_u\widetilde{Q}\,\,\,\,\,=\,\,\,\widetilde{Q}%
\,^A\,_{B;k}.u^k.\partial _A\otimes dx^B$, $\widetilde{Q}=\widetilde{Q}%
\,^A\,_B.\partial _A\otimes dx^B $, 
\[
\begin{array}{c}
\widetilde{Q}\,_{\,\,\,\,\,\,B;k}^A=(d_{\underline{K}})^\omega
.Q_{\,\,\,\,\,\,\,\,B;k}^A+[(d_{\underline{K}})^\omega
]_{;k}.Q^A\,_B\,+2\omega .U_{kl}^l.\widetilde{Q}\,^A\,_B= \\ 
=\widetilde{Q}\,_{\,\,\,\,\,\,\,B,k}^A+\Gamma _{Ck}^A.\widetilde{Q}%
\,_{\,\,\,\,\,\,B}^C+P_{Bk}^D.\widetilde{Q}\,^A\,_D+2\omega .P_{kl}^l.%
\widetilde{Q}\,^A\,_B.
\end{array}
\]
$\widetilde{Q}\,_{\,\,\,\,\,\,B;k}^A$ are the \textit{components of the
covariant derivative of type 2.} of the covariant tensor density $\widetilde{%
Q}$ \textit{in a co-ordinate basis.}

The properties of the form-invariant covariant differential operators are
determined by the properties of the covariant differential operator and
their construction.

\subsection{Necessary and sufficient conditions for the application of the
Stokes theorem as conditions on the affine connections}

By means of the different form-invariant covariant differential operators
different necessary and sufficient conditions can be found for the
application of the theorem of Stokes allowing the use of the boundary
conditions on the shell $(V_n)$ of the volume $V_n$.

(a) Under the action of $\overline{\nabla }_u\,$ the necessary and
sufficient conditions for the existence of the relation $\mathbf{j}^i\,_{;i}=%
\mathbf{j}^i$ $_{,i}$ for $\forall \,\,\mathbf{j}^i$ are the conditions 
\begin{equation}  \label{I-2.21}
\begin{array}{c}
\overline{q}_i=P_{ki}^k+\Gamma _{ik}^k=T_{ki}^k-C_{ik}\,^k+g_l^k.g_{k;i}^l=0%
\text{ (in a non-co-ordinate basis),} \\ 
\overline{q}_i=P_{ki}^k+\Gamma _{ik}^k=T_{ki}^k+g_l^k.g_{k;i}^l=0\text{ (in
a co-ordinate basis). }
\end{array}
\end{equation}

(b) Under the action of $\widetilde{\nabla }_u$ the necessary and sufficient
conditions for the existence of the relation $\mathbf{j}^i\,_{;i}=\mathbf{j}%
^i$ $_{,i}$ for $\forall \,\,\mathbf{j}^i$ are the conditions 
\begin{equation}  \label{I-2.22}
\begin{array}{c}
\widetilde{q}_i=P_{ik}^k+\Gamma _{ik}^k-C_{ik}\,^k=0\text{ \thinspace
\thinspace (in a non-co-ordinate basis),} \\ 
\widetilde{q}_i=P_{ik}^k+\Gamma _{ik}^k=0\,\,\,\,\text{(in a co-ordinate
basis). }
\end{array}
\end{equation}

Under the conditions in a co-ordinate basis the Stokes theorem can be
applied for the common divergency $\mathbf{j}^i\,_{,i}$ in $(\overline{L}%
_n,g)$-space $(\dim M=n)$.

\textit{Explicit form of the canonical Euler-Lagrange equations for} $g_{ij}$%
\begin{equation}  \label{A.10}
\delta _g\mathbf{L}=\frac{\delta _g\mathbf{L}}{\delta g_{kl}}.\delta
g_{kl}+\,_g\mathbf{j}^m\text{ }_{;m}\text{ ,}
\end{equation}
\begin{equation}  \label{A.11}
\frac{\delta _g\mathbf{L}}{\delta g_{kl}}=0\text{ ,}
\end{equation}
\begin{equation}  \label{A.12}
_g\mathbf{j}^m=[\frac{\partial \mathbf{L}}{\partial g_{kl;m}}-(\frac{%
\partial \mathbf{L}}{\partial g_{kl;m;n}}+\frac{\partial \mathbf{L}}{%
\partial g_{kl;n;m}})_{;n}].\delta g_{kl}+(\frac{\partial \mathbf{L}}{%
\partial g_{kl;n;m}}.\delta g_{kl})_{;n}\text{ .}
\end{equation}

\textit{Explicit form of the canonical Euler-Lagrange equations for} $V^A$ $%
_B$%
\begin{equation}  \label{A.13}
\delta _v\mathbf{L}=\frac{\delta _v\mathbf{L}}{\delta V^A\text{ }_B}.\delta
V^A\text{ }_B+\,_v\mathbf{j}^i\text{ }_{;i}\text{ ,}
\end{equation}
\begin{equation}  \label{A.14}
\frac{\delta _v\mathbf{L}}{\delta V^A\text{ }_B}=0\text{ ,}
\end{equation}
\begin{equation}  \label{A.15}
_v\mathbf{j}^i=[\frac{\partial \mathbf{L}}{\partial V^A\text{ }_{B;i}}-(%
\frac{\partial \mathbf{L}}{\partial V^A\text{ }_{B;i;k}}+\frac{\partial 
\mathbf{L}}{\partial V^A\text{ }_{B;k;i}})_{;k}].\delta V^A\text{ }_B+(\frac{%
\partial \mathbf{L}}{\partial V^A\text{ }_{B;k;i}}.\delta V^A\text{ }_B)_{;k}%
\text{ .}
\end{equation}

\textit{Remark.} For $(L_n,g)$-spaces and $\widetilde{\nabla }_u=\nabla
_u-2q.\Gamma _{ki}^i.u^k$ in a co-ordinate basis the relation $\mathbf{j}%
^i\,_{;i}$ $=\mathbf{j}^i\,_{,i}$ is fulfilled and no additional conditions
for $\Gamma =-P$ are needed (Lovelock and Rund 1975).

\subsection{Non-equivalence between the covariant and the canonical
Euler-Lagrange equations under the different conditions for the affine
connections}

The canonical Euler-Lagrange equations for $g_{ij}$ and $V^A\,_B$ can be
expressed by the use of the covariant Euler-Lagrange equations under the
different conditions for the affine connections. There exist two different
types of relations between both types of E-L equations, depending on the
different form-invariant covariant differential operators $\overline{\nabla }%
_u$ and $\widetilde{\nabla }_u$. Both operators determine different
conditions on the affine connections for the application of the theorem of
Stokes and along with it the use of the boundary conditions.

1. If the form-invariant covariant operator $\overline{\nabla }_u$ is used,
then the canonical E-L equations can be written in the forms respectively 
\begin{equation}  \label{I-2.27}
\frac{\delta _g\mathbf{L}}{\delta g_{ij}}=0\rightleftharpoons \frac{\delta
_gL}{\delta g_{ij}}=-\frac 12.L.g^{\overline{i}\overline{j}}+\,_g\overline{D}%
\,^{ij}\text{ ,}
\end{equation}
\begin{equation}  \label{I-2.28}
\frac{\delta _v\mathbf{L}}{\delta V^A\text{ }_B}=0\rightleftharpoons \frac{%
\delta _vL}{\delta V^A\text{ }_B}=\,_v\overline{D}_A\text{ }^B\text{ ,}
\end{equation}

\noindent with the following relations:

(a) Relations between the covariant and canonical Euler-Lagrange equations
for $g_{ij}$%
\begin{equation}  \label{A.16}
\frac{\delta _g\mathbf{L}}{\delta g_{kl}}=\sqrt{-d_g}.(\frac{\overline{%
\delta }_gL}{\delta g_{kl}}-\,_g\overline{D}\,^{kl})=\sqrt{-d_g}.(\frac{%
\delta _gL}{\delta g_{kl}}+\frac 12.L.g^{\overline{k}\overline{l}}-\,_g%
\overline{D}\,^{kl})\text{ ,}
\end{equation}
\begin{equation}  \label{A.17}
_g\overline{D}\,^{kl}=\frac 12[\frac{\partial L}{\partial g_{kl;m}}.Q_m-%
\frac{\partial L}{\partial g_{kl;m;n}}.(Q_{n;m}+\frac 12.Q_n.Q_m)-(\frac{%
\partial L}{\partial g_{kl;m;n}}+\frac{\partial L}{\partial g_{kl;n;m}}%
)_{;n}.Q_m]\text{ ,}
\end{equation}
\begin{equation}  \label{A.18}
Q_i=g^{\overline{k}\overline{l}}.g_{kl;i}=[\log (-d_g)]_{;i}\text{
,\thinspace \thinspace \thinspace \thinspace \thinspace \thinspace }_g%
\overline{D}\,^{kl}=\,_g\overline{D}\,^{lk}\text{ .}
\end{equation}

The relation between $_gj^i$ and $_g\mathbf{j}^i$ can be found in the form 
\begin{equation}  \label{A.19}
_g\mathbf{j}^i=\sqrt{-d_g}(_gj^i-\,_g\overline{D}\,^i)\text{ , }
\end{equation}
\begin{equation}  \label{A.20}
_g\overline{D}\,^i=\frac 12.\frac{\partial L}{\partial g_{kl;i;m}}%
.Q_m.\delta g_{kl}=\,_g\overline{D}\,^{kli}.\delta g_{kl}\text{ ,}
\end{equation}
\begin{equation}  \label{A.21}
_g\overline{D}\,^{kli}=\frac 12.\frac{\partial L}{\partial g_{kl;i;m}}.Q_m%
\text{ ,}
\end{equation}

\noindent and the identity for $_g\overline{D}\,^k$%
\begin{equation}
_g\overline{D}\,^k\text{ }_{;k}\equiv \frac 12.Q_k.(_gj^k-\,_g\overline{D}%
\,^k)-\,_g\overline{D}\,^{kl}.\delta g_{kl}  \label{A.22}
\end{equation}

\noindent is valid.

(b) Relation between the covariant and canonical Euler-Lagrange equations
for $V^A$ $_B$%
\begin{equation}  \label{A.23}
\frac{\delta _v\mathbf{L}}{\delta V^A\text{ }_B}=\sqrt{-d_g}(\frac{\delta _vL%
}{\delta V^A\text{ }_B}-\,_v\overline{D}_A\text{ }^B)\text{ ,}
\end{equation}
\begin{equation}  \label{A.24}
_v\overline{D}_A\text{ }^B=\frac 12[\frac{\partial L}{\partial V^A\text{ }%
_{B;i}}.Q_i-\frac{\partial L}{\partial V^A\text{ }_{B;i;k}}.(Q_{k;i}+\frac
12.Q_k.Q_i)-Q_i.(\frac{\partial L}{\partial V^A\text{ }_{B;i;k}}+\frac{%
\partial L}{\partial V^A\text{ }_{B;k;i}})_{;k}]\text{ .}
\end{equation}

The relation between $_vj^i$ and $_v\mathbf{j}^i$ can be expressed in the
form 
\begin{equation}  \label{A.25}
_v\mathbf{j}^i=\sqrt{-d_g}(_vj^i-\,_v\overline{D}\,^i)\text{ ,}
\end{equation}
\begin{equation}  \label{A.26}
_v\overline{D}\,^i=\frac 12.Q_k.\frac{\partial L}{\partial V^A\text{ }%
_{B;i;k}}.\delta V^A\text{ }_B\text{ ,}
\end{equation}

\noindent and the identity 
\begin{equation}
_v\overline{D}\,^k\text{ }_{;k}\equiv \frac 12.Q_k.(_vj^k-\,_v\overline{D}%
\,^k)-\,_v\overline{D}_A\text{ }^B.\delta V^A\text{ }_B  \label{A.27}
\end{equation}

\noindent is valid.

2. If the form-invariant covariant operator $\widetilde{\nabla }_u$ is used,
then the canonical E-L equations can be written in the forms respectively

\begin{equation}  \label{I-2.30}
\frac{\delta _g\mathbf{L}}{\delta g_{ij}}=0\rightleftharpoons \frac{\delta
_gL}{\delta g_{ij}}=-\frac 12.L.g^{\overline{i}\overline{j}}+\,_g\widetilde{D%
}\,^{ij}\text{ ,}
\end{equation}
\begin{equation}  \label{I-2.31}
\frac{\delta _v\mathbf{L}}{\delta V^A\text{ }_B}=0\rightleftharpoons \frac{%
\delta _vL}{\delta V^A\text{ }_B}=\,_v\widetilde{D}_A\text{ }^B\text{ ,}
\end{equation}

\noindent with the following relations:

(a) Relations between the covariant and canonical Euler-Lagrange equations
for $g_{ij}$%
\begin{equation}  \label{I-2.32}
\frac{\delta _g\mathbf{L}}{\delta g_{kl}}=\sqrt{-d_g}(\frac{\overline{\delta 
}_gL}{\delta g_{kl}}-\,_g\widetilde{D}\,^{kl})=\sqrt{-d_g}(\frac{\delta _gL}{%
\delta g_{kl}}+\frac 12.L.g^{\overline{k}\overline{l}}-\,_g\widetilde{D}%
\,^{kl})\text{ ,}
\end{equation}
\[
_g\widetilde{D}\,^{kl}=(\frac 12.Q_m-U_{rm}\,^r).\frac{\partial L}{\partial
g_{kl;m}}-(\frac 12.Q_n-U_{rn}\,^r)_{;m}.\frac{\partial L}{\partial
g_{kl;m;n}}- 
\]
\begin{equation}  \label{I-2.33}
-(\frac 12.Q_m-U_{rm}\,^r).[\frac 12(Q_n-U_{sn}\,^s).\frac{\partial L}{%
\partial g_{kl;m;n}}+(\frac{\partial L}{\partial g_{kl;m;n}}+\frac{\partial L%
}{\partial g_{kl;n;m}})_{;n}]=\,_g\widetilde{D}\,^{lk}\text{ ,}
\end{equation}
\begin{equation}  \label{I-2.34}
Q_i=g^{\overline{k}\overline{l}}.g_{kl;i}=[\log (-d_g)]_{;i}\text{ , }%
U_{li}\,^k=P_{li}^k-P_{il}^k\,\,\,(-C_{li}\,^k)\text{ ,\thinspace \thinspace 
}U_{rm}\,^r=g_k^r.U_{rm}\,^k\text{ .}
\end{equation}

The relation between $_gj^i$ and $_g\mathbf{j}^i$ can be found in the form 
\begin{equation}  \label{I-2.35}
_g\mathbf{j}^i=\sqrt{-d_g}(_gj^i-\,_g\widetilde{D}\,^i)\text{ ,}
\end{equation}
\begin{equation}  \label{I-2.36}
_g\widetilde{D}\,^i=(\frac 12.Q_m-U_{rm}\,^r).\frac{\partial L}{\partial
g_{kl;i;m}}.\delta g_{kl}=\,_g\widetilde{D}\,^{kli}.\delta g_{kl}\text{ ,}
\end{equation}
\begin{equation}  \label{I-2.37}
_g\widetilde{D}\,^{kli}=(\frac 12.Q_m-U_{rm}\,^r).\frac{\partial L}{\partial
g_{kl;i;m}}\text{ ,}
\end{equation}

\noindent and the identity for $_g\widetilde{D}\,^k$%
\begin{equation}
_g\widetilde{D}\,^k\text{ }_{;k}\equiv (\frac 12.Q_k-U_{rk}\,^r).(_gj^k-\,_g%
\widetilde{D}\,^k)-\,_g\widetilde{D}\,^{kl}.\delta g_{kl}  \label{I-2.38}
\end{equation}

\noindent is valid.

(b) Relation between the covariant and canonical Euler-Lagrange equations
for $V^A$ $_B$%
\begin{equation}  \label{I-2.39}
\frac{\delta _v\mathbf{L}}{\delta V^A\text{ }_B}=\sqrt{-d_g}(\frac{\delta _vL%
}{\delta V^A\text{ }_B}-\,_v\widetilde{D}_A\text{ }^B)\text{ ,}
\end{equation}
\[
_v\widetilde{D}_A\text{ }^B=(\frac 12.Q_i-U_{ri}\,^r).\frac{\partial L}{%
\partial V^A\text{ }_{B;i}}-(\frac 12.Q_k-U_{rk}\,^r)_{;i}.\frac{\partial L}{%
\partial V^A\text{ }_{B;i;k}}- 
\]
\begin{equation}  \label{I-2.40}
-(\frac 12.Q_i-U_{ri}\,^r).[(\frac 12.Q_k-U_{mk}\,^m).\frac{\partial L}{%
\partial V^A\,_{B;i;k}}+(\frac{\partial L}{\partial V^A\text{ }_{B;i;k}}+%
\frac{\partial L}{\partial V^A\text{ }_{B;k;i}})_{;k}]\text{ .}
\end{equation}

The relation between $_vj^i$ and $_v\mathbf{j}^i$ can be expressed in the
form 
\begin{equation}  \label{I-2.41}
_v\mathbf{j}^i=\sqrt{-d_g}(_vj^i-\,_v\widetilde{D}\,^i)\text{ ,}
\end{equation}
\begin{equation}  \label{I-2.42}
_v\widetilde{D}\,^i=(\frac 12.Q_k-U_{mk}\,^m).\frac{\partial L}{\partial V^A%
\text{ }_{B;i;k}}.\delta V^A\text{ }_B\text{ ,}
\end{equation}

\noindent and the identity 
\begin{equation}
_v\widetilde{D}\,^k\text{ }_{;k}\equiv (\frac
12.Q_k-U_{mk}\,^m).(_vj^k-\,_vD^k)-\,_v\widetilde{D}_A\text{ }^B.\delta V^A%
\text{ }_B  \label{I-2.43}
\end{equation}

\noindent is valid.

It is obvious (compare for instance \ref{I-2.19} and \ref{I-2.20} with \ref
{I-2.8} for the case of $\overline{\nabla }_u$) that in $(\overline{L}_n,g)$
-spaces the covariant Euler-Lagrange equations are different from the
canonical Euler-Lagrange equations if the explicit form of $j^i$ and $%
\mathbf{j}^i$ are not taken into account and only general conditions for
every $j^i$ or for every $\mathbf{j}^i$ are imposed . The difference is a
corollary of using different terms of $\delta \mathbf{L}$ and different
conditions for $\Gamma $ and $P$ in both ways for obtaining common
divergency term necessary for applying the Stokes theorem. Both type of
equations are identical for $V_n$ and $U_n$-spaces, where $g_{ij;k}=0$, $%
(Q_i=g^{kl}.g_{kl;i}=0)$ and therefore $_g\overline{D}\,^{ij}=0$ and $_v%
\overline{D}_A$ $^B=0$. In $(\overline{L}_n,g)$-spaces they would be
equivalent in a coordinate basis, if $g^{\overline{k}\overline{l}%
}.g_{kl;i}=0 $. On the other side, it is obviously that, because of the
different conditions for $\Gamma $ and $P$, in a $(\overline{L}_n,g)$-space
in general \textit{only one type} of Euler-Lagrange's equations (canonical
or covariant) could be used. Therefore, \textit{the unique way for avoiding
ambiguities is the use of the covariant Euler-Lagrange equations without
using additional conditions on the affine connections}.

\textit{Remark.} The dilemma of choosing the right type of the
Euler-Lagrange equations can be easily solved in the case of $(L_n,g)$%
-spaces by using only the canonical E-L equations and the definition of $%
\widetilde{\nabla }_u$ (which in this case requires no additional conditions
for the affine connection $\Gamma =-P$).

\section{Energy-momentum tensors}

The energy-momentum tensors can be determined using the method of
Lagrangians with covariant derivatives which leads to the covariant Noether
identities.

\subsection{Lie variation of the Lagrangian density $\mathbf{L}$}

Let a Lagrangian density of the type \ref{1.2} 
\[
\mathbf{L}=\sqrt{-d_g}.L(g_{ij}\text{, }g_{ij;k}\text{, }g_{ij;k;l}\text{, }%
V^A\text{ }_B\text{, }V^A\text{ }_{B;i}\text{, }V^A\text{ }_{B;i;j})\text{ } 
\]

\noindent be given. The action of the system is defined as 
\begin{equation}
S=\dint\limits_{V_n}\mathbf{L.}d^{(n)}x=\dint\limits_{V_n}L.d\omega \text{ ,
\thinspace }  \label{II-3.0}
\end{equation}

\noindent where 
\[
\begin{array}{c}
\text{\thinspace \thinspace }L.d\omega =\mathbf{L}.\frac 1{n!}.\varepsilon
_A.\omega ^A=\mathbf{L}.d^{(n)}x\text{ , }d\omega =\sqrt{-d_g}%
.d^{(n)}x=\frac 1{n!}.\sqrt{-d_g}.\varepsilon _A.\omega ^A\text{ ,} \\ 
\varepsilon _A=\varepsilon _{i_1...i_n}\text{ , }\omega ^A=dx^{i_1}\wedge
...\wedge dx^{i_n}\text{ , }d^{(n)}x=dx^1\wedge ...\wedge dx^n\text{
,\thinspace }\dim M=n\text{ .}
\end{array}
\]

$\varepsilon _A$ is the full antisymmetric Levi-Civita symbol (Lovelock and
Rund 1975). The Lie derivative of $S$ equal to the Lie variation of $S$ will
have the form 
\begin{equation}  \label{II-3.0a}
\begin{array}{c}
\pounds _\xi S=\pounds _\xi [\dint\limits_{V_n} \mathbf{L.}d^{(n)}x]=\pounds
_\xi [\dint\limits_{V_n}L.d\omega ]=\dint\limits_{V_n}\pounds _\xi [\mathbf{%
L.}d^{(n)}x]=\dint\limits_{V_n}\pounds _\xi [L.d\omega ]= \\ 
=\dint\limits_{V_n}(\overline{\pounds }_\xi \mathbf{L).}d^{(n)}x\text{ .}
\end{array}
\end{equation}

$\overline{\pounds }_\xi \mathbf{L}$ is the Lie derivative of the Lagrangian
density $\mathbf{L}$ obtained as a result of the action of the
form-invariant Lie differential operator $\overline{\pounds }_\xi $ on a
tensor density of the type of $\mathbf{L}$. Let we now make some comments on
the definitions of form-invariant Lie differential operators acting on
tensor densities over $(\overline{L}_n,g)$-spaces.

\subsection{Form-invariant Lie differential operators for tensor densities}

The determination of Lie differential operators acting on a tensor density
and preserving its type and weight, i. e. mapping a density in a tensor
density of the same type, is possible but not unique.

\textbf{Lie derivatives of tensor densities}. The Lie differential operator
acts on contravariant and covariant tensor densities in accordance with its
action on functions and tensor fields over a manifold $M$:$\,$%
\[
\begin{array}{c}
\pounds _\xi \overline{Q}=(d_{\overline{K}})^\omega .\pounds _\xi Q+\omega
.[\nabla _\xi (\log d_{\overline{K}})].\overline{Q}-2\omega .\Gamma
_{ik}^i.\xi ^k.\overline{Q}\,\,\,\text{,}\, \\ 
\,\pounds _\xi \widetilde{Q}=(d_{\underline{K}})^\omega .\pounds _\xi
Q+\omega .[\nabla _\xi \log d_{\underline{K}}].Q-2\omega .P_{\beta \gamma
}^\beta .\xi ^\gamma .\widetilde{Q}\text{ .}
\end{array}
\]
The transformation properties of $\pounds _\xi \overline{Q}$ and $\pounds
_\xi \widetilde{Q}$ are determined by the construction of $\overline{Q}$ and 
$\widetilde{Q}$, their transformation properties and the transformation
properties of the derivatives of the Jacobian $J$. From $\overline{Q}%
^{\prime }=J^{-2\omega }.\overline{Q}=J^{-2\omega }.(d_{\overline{K}%
})^\omega .Q$, $\omega =q\neq 0$, it follows that $\pounds _\xi \overline{Q}%
^{\prime }=-2\omega .[\xi (\log J)].\overline{Q}^{\prime }+J^{-2\omega
}.\pounds _\xi \overline{Q}$. It is obvious that $\pounds _\xi \overline{Q}$
does not transform as a tensor density, i. e. the condition $\pounds _\xi 
\overline{Q}^{\prime }=J^{-2\omega }.\pounds _\xi \overline{Q}$ is not
fulfilled in general. From $\widetilde{Q}^{\prime }=J^{2\omega }.\widetilde{Q%
}=J^{2\omega }.(d_{\underline{K}})^\omega .Q=\widetilde{Q}^{A^{\prime
}}{}_{B^{\prime }}.e_{A^{\prime }}\otimes e^{B^{\prime }}=J^{2\omega }.%
\widetilde{Q}^A{}_B.e_A\otimes e^B$, it follows that $\pounds _\xi 
\widetilde{Q}^{\prime }=2\omega .[\xi (\log J)].\widetilde{Q}^{\prime
}+J^{2\omega }.\pounds _\xi \widetilde{Q}$, and that $\pounds _\xi 
\widetilde{Q}$ does not transform as a covariant tensor density. It is
necessary a new Lie differential operator to be constructed which action on
a tensor density results again in a tensor density with the corresponding
transformation properties.

\textbf{Definition 10}. \textit{Form-invariant Lie differential operator of
type 1. for contravariant tensor densities }of the type of $\overline{Q}.$
Lie differential operator $\overline{\pounds }_\xi :$ $\overline{\pounds }%
_\xi =\pounds _\xi +2\omega \Gamma _{ik}^i.\xi ^k-2\omega (\xi ^i$ $%
_{;i}-T_{ik}^i.\xi ^k)$ (in a co-ordinate basis).

The result of the action of the form-invariant Lie differential operator on
a tensor density of the type of $\overline{Q}$ is called \textit{Lie
derivative of type 1. of the tensor density }$\overline{Q}$ along the
contravariant vector field $\xi :\overline{\pounds }_\xi \overline{Q}=%
\overline{\pounds }_\xi [(d_{\overline{K}})^\omega .Q]=[\pounds _\xi (d_{%
\overline{K}})^\omega ].Q+(d_{\overline{K}})^\omega .\pounds _\xi Q=%
\overline{\pounds }_\xi \overline{Q}\,^A$ $_B.\partial _A\otimes dx$, 
\[
\begin{array}{c}
\overline{\pounds }_\xi \overline{Q}\,^A\text{ }_B=\overline{Q}\,^A\text{ }%
_{B;k}.\xi ^k+(S_{Ci}\text{ }^{Aj}.\overline{Q}\,^C\text{ }_B-S_{Bi}\text{ }%
^{Dj}.\overline{Q}\,^A\text{ }_D).(\xi ^i\text{ }_{;j}-T_{jk}^i.\xi ^k)- \\ 
-2\omega .(\xi ^i\text{ }_{;i}-T_{ik}^i.\xi ^k).\overline{Q}\,^A\text{ }_B%
\text{.}
\end{array}
\]

$\overline{\pounds }_\xi \overline{Q}\,^A$ $_B$ are called \textit{%
components of the Lie derivative of type 1.} of the tensor density $%
\overline{Q}$ \textit{in a co-ordinate basis}.

\textbf{Definition 11}. \textit{Form-invariant Lie differential operator of
type 2. for contravariant tensor densities }of the type of $\overline{Q}.$
Lie differential operator $\widetilde{\pounds }_\xi :\widetilde{\pounds }%
_\xi =\pounds _\xi +2\omega .\Gamma _{ik}^i.\xi ^k-2\omega .\xi ^i$ $_{;i}$
(in a co-ordinate basis).

The result of the action of the form-invariant Lie differential operator $%
\widetilde{\pounds }_\xi $ on a tensor density $\overline{Q}$ is called 
\textit{Lie derivative of type 2. of the contravariant tensor density }$%
\overline{Q}$ along the contravariant vector field $\xi :\widetilde{\pounds }%
_\xi \overline{Q}=\widetilde{\pounds }_\xi [(d_{\overline{K}})^\omega
.Q]=(d_{\overline{K}})^\omega .\pounds _\xi Q+[\pounds _\xi (d_{\overline{K}%
})^\omega ].Q-2\omega .T_{ik}^i.\xi ^k.\overline{Q}=\widetilde{\pounds }_\xi 
\overline{Q}\,^A$ $_B.\partial _A\otimes dx^B$, 
\[
\begin{array}{c}
\widetilde{\pounds }_\xi \overline{Q}\,^A\text{ }_B=\overline{Q}\,^A\text{ }%
_{B;k}.\xi ^k+(S_{Ci}\text{ }^{Aj}.\overline{Q}\,^C\text{ }_B-S_{Bi}\text{ }%
^{Dj}.\overline{Q}\,^A\text{ }_D).(\xi ^i\text{ }_{;j}-T_{jk}^i.\xi ^k)- \\ 
-2\omega .\xi ^i\text{ }_{;i}.\overline{Q}\,^A\text{ }_B\text{.}
\end{array}
\]

$\widetilde{\pounds }_\xi \overline{Q}\,^A$ $_B$ are called \textit{%
components of the Lie derivative of type 2. of the tensor density }$%
\overline{Q}$ in a co-ordinate basis.

\textbf{Definition 12. }\textit{Form-invariant Lie differential operator of
type 1. for covariant tensor densities }of the type of $\widetilde{Q}$. Lie
differential operator $\overline{\pounds }_\xi :$ $\overline{\pounds }_\xi
=\pounds _\xi +2\omega .P_{ik}^i.\xi ^k+2\omega .(\xi ^i$ $%
_{;i}-T_{ik}^i.\xi ^k)$ (in a co-ordinate basis).

The result of the action of the form-invariant differential operator $%
\overline{\pounds }_\xi $ on a tensor density $\widetilde{Q}$ is called 
\textit{Lie derivative of type 1. of the covariant tensor density }$%
\widetilde{Q}$ along a contravariant vector field $\xi :\overline{\pounds }%
_\xi \widetilde{Q}=\overline{\pounds }_\xi [(d_{\underline{K}})^\omega
.Q]=[\pounds _\xi (d_{\underline{K}})^\omega ].Q+(d_{\underline{K}})^\omega
.\pounds _\xi Q=\overline{\pounds }_\xi \widetilde{Q}\,^A$ $_B.\partial
_A\otimes dx^B$. 
\[
\begin{array}{c}
\overline{\pounds }_\xi \widetilde{Q}\,^A\text{ }_B=\widetilde{Q}\,^A\text{ }%
_{B;k}.\xi ^k+(S_{Ci}\text{ }^{Aj}.\widetilde{Q}\,^C\text{ }_B-S_{B\overline{%
i}}\text{ }^{D\underline{j}}.\widetilde{Q}\,^A\text{ }_D).(\xi ^i\text{ }%
_{;j}-T_{jk}^i.\xi ^k)+ \\ 
+2\omega .(\xi ^i\text{ }_{;i}-T_{ik}^i.\xi ^k).\widetilde{Q}\,^A\text{ }_B%
\text{.}
\end{array}
\]

$\overline{\pounds }_\xi \widetilde{Q}\,^A$ $_B$ are called \textit{%
components of the Lie derivative of type 1. of the tensor density }$%
\widetilde{Q}$ in a co-ordinate basis.

\textbf{Definition 13}. \textit{Form-invariant Lie differential operator of
type 2. for covariant tensor densities }of the type $\widetilde{Q}.$ Lie
differential operator $\widetilde{\pounds }_\xi :\widetilde{\pounds }_\xi
=\pounds _\xi +2\omega .P_{ik}^i.\xi ^k+2\omega .\xi ^i$ $_{;i}=\pounds _\xi
+2\omega .P_{ki}^i\xi ^k+2\omega .(\xi ^k\,_{;k}+U_{ik}^i.\xi ^k)$ (in a
co-ordinate basis).

The result of the action of the form-invariant differential operator $%
\widetilde{\pounds }_\xi $ on a tensor density $\widetilde{Q}$ is called 
\textit{Lie derivative of type 2. of the covariant tensor density }$%
\widetilde{Q}$ along a contravariant vector field $\xi :\widetilde{\pounds }%
_\xi \widetilde{Q}=\widetilde{\pounds }_\xi [(d_{\underline{K}})^\omega
.Q]=[\pounds _\xi (d_{\underline{K}})^\omega ].Q+(d_{\underline{K}})^\omega
.\pounds _\xi Q+2\omega .T_{ik}^i.\xi ^k.\widetilde{Q}=\widetilde{\pounds }%
_\xi \widetilde{Q}\,^A$ $_B.\partial _A\otimes dx^B$, 
\[
\begin{array}{c}
\widetilde{\pounds }_\xi \widetilde{Q}\,^A\text{ }_B=\widetilde{Q}\,^A\text{ 
}_{B;k}.\xi ^k+(S_{Ci}\text{ }^{Aj}.\widetilde{Q}\,^C\text{ }_B-S_{B%
\overline{i}}\text{ }^{D\underline{j}}.\widetilde{Q}\,^A\text{ }_D).(\xi ^i%
\text{ }_{;j}-T_{jk}^i.\xi ^k)+ \\ 
+2\omega .\xi ^i\text{ }_{;i}.\widetilde{Q}\,^A\text{ }_B\text{.}
\end{array}
\]

$\widetilde{\pounds }_\xi \widetilde{Q}^A$ $_B$ are called \textit{%
components of the Lie derivative of type 2. of the tensor density }$%
\widetilde{Q}$ in a co-ordinate basis [compare with (Lovelock and Rund 1975
- p.124)].

It follows from \ref{II-3.0a} that \textit{the Lie variation of a Lagrangian
density }$\mathbf{L}$\textit{\ is related to the Lie derivative of type 1.
of the tensor density }$\mathbf{L}$\textit{\ }of the type of $\widetilde{Q}$
(rank $\widetilde{Q}=0$, \underline{$K$}$\,=g$, $\omega =q=\frac 12$).

\subsection{Covariant Noether's identities}

By means of the Lie variation of $\mathbf{L}$%
\[
\overline{\pounds }_\xi \mathbf{L}=\sqrt{-d_g}(L_{,i}.\xi ^i+\frac 12.L.g^{%
\overline{i}\overline{k}}.\pounds _\xi g_{ik})\equiv 
\]
\[
\equiv \frac{\partial \mathbf{L}}{\partial g_{ik}}.\pounds _\xi g_{ik}+\frac{%
\partial \mathbf{L}}{\partial g_{ik;l}}.\pounds _\xi (g_{ik;l})+\frac{%
\partial \mathbf{L}}{\partial g_{ik;l;m}}.\pounds _\xi (g_{ik;l;m})+ 
\]
\begin{equation}  \label{II-3.1}
+\frac{\partial \mathbf{L}}{\partial V^A\text{ }_B}.\pounds _\xi V^A\text{ }%
_B+\frac{\partial \mathbf{L}}{\partial V^A\text{ }_{B;i}}.\pounds _\xi (V^A%
\text{ }_{B;i})+\frac{\partial \mathbf{L}}{\partial V^A\text{ }_{B;i;j}}%
.\pounds _\xi (V^A\text{ }_{B;i;j})
\end{equation}

\noindent with an arbitrary contravariant vector field $\xi =\xi ^i.\partial
_i=\xi ^k.e_k$ and the explicit form of the Lie derivatives of the metric
and non-metric tensor field's components and their covariant derivatives (s.
Appendix) the Lie identity can be written in the form 
\begin{equation}
\overline{P}=(\overline{P}_i+\overline{P}_k\,^j.T_{ij}^k).\xi ^i+\overline{P}%
_i\,^j.\xi ^i\,_{;j}\equiv 0\text{ for }\forall \,\,\xi ^i\text{ and }%
\forall \,\xi ^i\,_{;j}\text{ ,}  \label{II-3.1a}
\end{equation}

\noindent or as the two identities $\overline{P}_i\equiv \overline{F}_i+%
\overline{\theta }_i$ $^j$ $_{;j}\equiv 0$ and $\overline{P}_i\,^j\equiv 
\overline{\theta }_i$ $^j-\,_sT_i$ $^j-\overline{Q}_i$ $^j\equiv 0$.

The covariant Noether identities can be found in (an analogous to the case
of $U_n$- and $(L_n,g)$-spaces (Manoff 1987, 1991) form 
\begin{equation}  \label{II-3.2}
\overline{F}_i+\overline{\theta }_i\text{ }^j\text{ }_{;j}\equiv
0\,\,\,\,\,\,\text{ \thinspace \thinspace \thinspace \thinspace \thinspace
\thinspace \thinspace \thinspace \thinspace \thinspace \thinspace \thinspace
\thinspace \thinspace \thinspace \thinspace \thinspace \thinspace \thinspace
\thinspace \thinspace (first covariant Noether's identity),}
\end{equation}
\begin{equation}  \label{II-3.3}
\overline{\theta }_i\text{ }^j-\,_sT_i\text{ }^j\equiv \overline{Q}_i\text{ }%
^j\text{ \thinspace \thinspace \thinspace \thinspace \thinspace \thinspace
\thinspace \thinspace \thinspace (second covariant Noether's identity),}
\end{equation}

\noindent where 
\begin{equation}
\begin{array}{c}
\overline{F}_i=\,_v\overline{F}_i+\,_g\overline{F}_i\text{ ,} \\ 
\overline{\theta }_i\text{ }^j=\,_v\overline{\theta }_i\text{ }^j+\,_g%
\overline{\theta }_i\text{ }^j\text{ ,} \\ 
_sT_i\text{ }^j=\,_{vs}T_i\text{ }^j+\,_{gs}T_i\text{ }^j\text{ ,} \\ 
\overline{Q}_i\text{ }^j=\,_v\overline{Q}_i\text{ }^j+\,_g\overline{Q}_i%
\text{ }^j\text{ .}
\end{array}
\label{II-3.4}
\end{equation}

The quantities $_vA$, $(A\sim \overline{F}_i$, $\overline{\theta }_i$ $^j$, $%
_sT_i$ $^j$, $\overline{Q}_i$ $^j)$ are the corresponding quantities to the
components $V^A$ $_B$ of the non-metric tensors $V$. The quantities $_gA$
are the corresponding quantities to the components $g_{ij}$ of the metric
tensor field $g$.

\subsection{Canonical, generalized canonical, symmetric (of Belinfante) and
variational (of Euler-Lagrange) energy-momentum tensors}

$\overline{\theta }_i$ $^j$ is the \textit{generalized canonical} \textit{%
energy-momentum tensor} (GC-EMT) constructed as a sum of the GC-EMT $_v%
\overline{\theta }_i$ $^j$ of the non-metric tensor fields $V$ and the
GC-EMT $_g\overline{\theta }_i$ $^j$ of the metric tensor $g$. 
\begin{equation}  \label{II-3.8}
\begin{array}{c}
_v \overline{\theta }_i\text{ }^j=\,_v\overline{t}_i\text{ }^j-\,_vK_i\text{ 
}^j-\,_v\overline{W}_i\text{ }^{jk}\text{ }_k\text{ ,} \\ 
_g\overline{\theta }_i\text{ }^j=\,_g\overline{t}_i\text{ }^j-\,_gK_i\text{ }%
^j-\,_g\overline{W}_i\text{ }^{jk}\text{ }_k\text{ .}
\end{array}
\end{equation}

$_v\overline{t}_i$ $^j$ and $_g\overline{t}_i$ $^j$ are the \textit{canonical%
} \textit{energy-momentum tensors} (C-EMT-s) of $V$ and $g$ correspondingly.
When $L$ is depending only on $V^A$ $_B$, $g_{ij}$ and their first \textit{%
partial }derivatives, $\overline{t}_i$ $^j$ has been symmetrized by means of
the a priory introduced s.c. Belinfante terms to the symmetric EMT of
Belinfante (Schmutzer 1968). Here $_v\overline{W}_i$ $^{jk}$ $_k$ and $_g%
\overline{W}_i$ $^{jk}$ $_k$ consist of the generalized Belinfante terms for
the non-metric tensor fields $V$ and for the metric $g$ respectively. They
appear in the construction of $\overline{\theta }_i$ $^j$ in a very natural
way as a result of a symmetrization procedure.

$_sT_i$ $^j$ is the \textit{symmetric energy-momentum tensor of Belinfante}
(S-EMT-B) constructed as a sum of the S-EMT-B $_{vs}T_i$ $^j$ for the
non-metric tensor fields $V$ and the S-EMT-B $_{gs}T_i$ $^j$ for the metric $%
g$, where (s. Appendix) 
\begin{equation}  \label{II-3.10}
\begin{array}{c}
_{vs}T_i \text{ }^j=\,_v\mathit{T}_i\text{ }^j-g_i^j.L\text{ ,} \\ 
_{gs}T_i\text{ }^j=\,_g\mathit{T}_i\text{ }^j-g_i^j.L\text{ .}
\end{array}
\end{equation}

$\overline{Q}_i$ $^j$ is the \textit{variational energy-momentum tensor of
Euler-Lagrange} (V-EMT-EL) constructed as a sum of the V-EMT-EL $_v\overline{%
Q}_i$ $^j$ for the tensor fields $V$ and the V-EMT-EL $_g\overline{Q}_i$ $^j$
for the metric $g$.

If the covariant Euler-Lagrange equations (E-L equations) for $V^A$ $_B$ are
of the type 
\begin{equation}  \label{II-3.11a}
\frac{\delta _vL}{\delta V^A\text{ }_B}=0\,\,\text{,}\,\,\,\,\text{%
and\thinspace \thinspace \thinspace \thinspace \thinspace \thinspace }\frac{%
\delta _gL}{\delta g_{ij}}=0\,\,\,\,\text{for\thinspace \thinspace
\thinspace }g_{ij},
\end{equation}

\noindent then $\overline{Q}_i$ $^j=0$ and 
\begin{equation}
\overline{\theta }_i\text{ }^j=\,_sT_i\text{ }^j\text{ .}  \label{II-3.12}
\end{equation}

But one has to take into account that the covariant E-L equations for the
metric $g$ and the non-metric tensor fields $V$ in $(\overline{L}_n,g)$%
-spaces are not the equations in \ref{II-3.11a} but the equations 
\begin{equation}  \label{II-3.12a}
\frac{\delta _gL}{\delta g_{ij}}=-\frac 12.L.g^{\overline{i}\overline{j}%
}-P^{ij}\text{ ,\thinspace \thinspace \thinspace \thinspace \thinspace
\thinspace \thinspace \thinspace \thinspace \thinspace \thinspace \thinspace
\thinspace }\frac{\delta _vL}{\delta V^A\,_B}=-P_A\,^B\text{ .}
\end{equation}

\textit{Remark}. The explicit forms of the EMT-s and their corresponding
quantities are given in the Appendix. The explicit form of $\overline{Q}_i$ $%
^j$ shows the role of the covariant Euler-Lagrange equations in its
structure.

\subsection{Symmetric energy-momentum tensor of Hilbert}

The \textit{symmetric energy-momentum tensor of Hilbert} (S-EMT-H) for the
components of non-metric tensor fields $V$ is determined in $V_n$-spaces
(where $S=C$ and $f^i\,_j=g_j^i$, $g_{\overline{i}k}=g_{ik}$, $g_{\underline{%
j}k}=g_{jk}$, $_{vsh}T^{\overline{k}j}=\,_{vsh}T^{kj}$ ) as (Schmutzer 1968) 
\begin{equation}  \label{II-3.13}
_{vsh}T_i\text{ }^j=-\frac 2{\sqrt{-d_g}}.\frac{\delta _g\mathbf{L}}{\delta
g_{\underline{j}k}}.g_{\overline{i}k}=g_{\overline{i}k}._{vsh}T^{\overline{k}%
j}\text{ .}
\end{equation}

The same definition could be accepted for $_{vsh}T_i$ $^j$ in $(L_n,g)$- and 
$(\overline{L}_n,g)$-spaces as a generalizations of that one in $V_n$-spaces.

By means of the explicit form of $\delta _g\mathbf{L}/\delta g_{jk}$ and the
form-invariant covariant differential operator $\overline{\nabla }_u$, $%
_{vsh}T_i$ $^j$ can also be written in the forms 
\begin{equation}  \label{II-3.14}
\begin{array}{c}
_{vsh}T_i \text{ }^j=\,_g\overline{Q}_i\text{ }^j-g_i^j.L+2._g\overline{D}%
\,^{\underline{j}k}.g_{\overline{i}k}= \\ 
=\,_{gs}T_i \text{ }^j-\,_g\mathit{T}_i\text{ }^j+\,_g\overline{Q}_i\text{ }%
^j+2._g\overline{D}\,^{\underline{j}k}.g_{\overline{i}k}= \\ 
=\,_{vs}T_i\text{ }^j-\,_v\mathit{T}_i\text{ }^j+\,_g\overline{Q}_i\text{ }%
^j+2._g\overline{D}\,^{\underline{j}k}.g_{\overline{i}k}\text{ .}
\end{array}
\end{equation}

The last two relations are obtained using the connection between $g_i^j.L$
and $_{vs}T_i$ $^j$ or $_{gs}T_i^{.j}$ in \ref{II-3.10}.

\textbf{Non-equivalence between the symmetric energy-momentum tensor of
Belinfante and the symmetric energy-momentum tensor of Hilbert}

It is obvious from \ref{II-3.14} that there is in general no equivalence
between the S-EMT-H and the S-EMT-B. The connection between $%
\,\,\,\,\,_{vs}T_i\,^j$ and $_{vsh}T_i$ $^j$ is based only on the common for
both tensors term $g_i^j.L$. The same is valid to the connection between $%
_{gs}T_i\,^j$ and $_{vsh}T_i$ $^j$. By means of \ref{II-3.14} one can easily
prove the following proposition:

\textbf{Proposition 2}. The necessary and sufficient conditions for the
equivalence between the S-EMT-B and the S-EMT-H ($_{vs}T_i{}^j=\,_{vsh}T_i%
\,^j$) are the conditions 
\begin{equation}  \label{II-3.17}
_v\mathit{T}_i\,^j=\,_g\overline{Q}_i\text{ }^j+2._g\overline{D}\,^{%
\underline{j}k}.g_{\overline{i}k}\text{ .}
\end{equation}

Therefore, $_{vs}T_i$ $^j$ is not equal in general to $_{vsh}T_i$ $^j$. This
fact has to be taken into account when energy-momentum problems are
considered in Lagrangian theories for tensor fields in $(\overline{L}_n,g)$%
-spaces.

\textbf{Non-equivalence between the generalized canonical energy}$-$\textbf{%
momentum tensor and the symmetric energy-momentum tensor of Hilbert}

The second covariant Noether identity $\overline{\theta }_i$ $^j-\,_sT_i$ $%
^j\equiv \overline{Q}_i$ $^j$,

\noindent written in the form 
\begin{equation}
_v\overline{\theta }_i\text{ }^j+\,_g\overline{\theta }_i\text{ }^j-(_{vs}T_i%
\text{ }^j+\,_{gs}T_i\text{ }^j)\equiv \,_v\overline{Q}_i\text{ }^j+\,_g%
\overline{Q}_i\text{ }^j\text{ ,}  \label{II-3.18}
\end{equation}

\noindent can be expressed by means of the relation 
\[
_g\overline{Q}_i\text{ }^j=\,_{vsh}T_i\text{ }^j+g_i^j.L-2._g\overline{D}\,^{%
\underline{j}k}.g_{\overline{i}k} 
\]
in the form 
\begin{equation}
_v\overline{\theta }_i\text{ }^j+\,_g\overline{\theta }_i\text{ }^j-(_{vs}T_i%
\text{ }^j+\,_{gs}T_i\text{ }^j)\equiv \,_v\overline{Q}_i\text{ }%
^j+\,_{vsh}T_i\text{ }^j+g_i^j.L-2._g\overline{D}\,^{\underline{j}k}.g_{%
\overline{i}k}\text{ .}  \label{II-3.20}
\end{equation}

If the covariant E-L equations for the non-metric tensor fields \ref
{II-3.11a} are fulfilled, then $_v\overline{Q}_i\,^j=0$ and 
\begin{equation}  \label{II-3.21}
_v\overline{\theta }_i\text{ }^j+\,_g\overline{\theta }_i\text{ }^j-(_v%
\mathit{T}_i\text{ }^j+\,_{gs}T_i\text{ }^j)=\,_{vsh}T_i\text{ }^j-2._g%
\overline{D}\,\,^{\underline{j}k}.g_{\overline{i}k}\text{ . }
\end{equation}

It is obvious that there is in general no equivalence between the GC-EMT and
the S-EMT-H. From the last equality it follows immediately the proposition:

\textbf{Proposition 3}. The necessary and sufficient conditions for the
equivalence between the GC-EMT and the S-EMT-H ($_v\overline{\theta }%
_i\,^j=\,_{vsh}T_i\,^j$) are the conditions 
\begin{equation}  \label{II-3.22}
_v\mathit{T}_i\text{ }^j=\,\,_g\overline{\theta }_i\text{ }^j-\,_{gs}T_i%
\text{ }^j+2._g\overline{D}\,\,^{\underline{j}k}.g_{\overline{i}k}\text{ .}
\end{equation}

These conditions are not fulfilled in general.

\section{Conclusions}

It has been shown that the corresponding to a Lagrangian density covariant
Noether identities contain in their structures the structure of the
covariant Euler-Lagrange equations. The Euler-Lagrange equations, the
covariant Noether identities and their corresponding energy-momentum tensors
built a full theoretical scheme of a Lagrangian theory of tensor fields with
compatible structure's elements.

The covariant Noether identities induce the determination of three types of
energy-momentum tensors: generalized canonical, symmetric (of Belinfante)
and variational (of Euler-Lagrange). The last type of EMT-s for non-metric
tensor fields $V$ vanishes in $V_n$-spaces [but not in $(\overline{L}_n,g)$%
-spaces] if the covariant Euler-Lagrange equations for the corresponding
dynamical fields are fulfilled. It does not vanish for the covariant metric
tensor field $g$. \textit{The different types of EMT-s appear for every
field variable (tensor field) in a Lagrangian density regardless of the role
of these field variables as dynamical or non-dynamical tensor fields}. The
symmetric energy-momentum tensor of Hilbert introduced in the Einstein field
equations appears as an irrelevant element in the whole scheme of a
covariant Lagrangian theory of tensor fields although some connections
between it and the other EMT-s could be established. It has been shown that
the Euler-Lagrange equations can be found in an independent of the affine
connections way and therefore, in an independent of the type of transport of
the tensor field manner.

\appendix

\section{Explicit form of the energy-momentum tensors}

\subsection{Lie derivatives of components and their covariant derivatives of
tensor fields}

1. Lie derivatives of the components $V^A\,_B$ of non-metric tensor fields $%
V $%
\begin{equation}  \label{a.1}
\begin{array}{c}
\pounds _\xi (V^A\,_B)=\pounds _\xi
V^A\,_B=[V^A\,_{B;l}+(S_{Ci}\,^{Ak}.V^C\,_B-S_{B \overline{i}}\,^{D%
\underline{k}}.V^A\,_D).T_{lk}\,^i].\xi ^l+ \\ 
+\,(S_{Ci}\,^{Aj}.V^C\,_B-S_{B\overline{i}}\,^{D\underline{j}}.V^A\,_D).\xi
^i\,_{;j}\,\,\,\,\,\,\text{, \thinspace \thinspace \thinspace \thinspace }%
T_{lk\,}\,^i\equiv T_{lk}^i\text{ .}
\end{array}
\end{equation}

2. Lie derivatives of the covariant derivatives of the components $V^A\,_B$
of non-metric fields $V$%
\begin{equation}  \label{a.2}
\begin{array}{c}
\pounds _\xi (V^A\,_{B;i})=[V^A\,_{B;i;j}+(S_{Cl}\,^{Ak}.V^C\,_{B;i}-S_{B 
\overline{l}}\,^{D\underline{k}}.V^A\,_{D;i}).T_{jk}^l+V^A%
\,_{B;l}.T_{ji}^l].\xi ^j+ \\ 
+(S_{Cl}\,^{Ak}.V^C\,_{B;i}-S_{B\overline{l}}\,^{D\underline{k}%
}.V^A\,_{D;i}).\xi ^l\,_{;k}+V^A\,_{B;l}.\xi ^l\,_{;i}\text{ ,}
\end{array}
\end{equation}
\begin{equation}  \label{a.3}
\begin{array}{c}
\pounds _\xi
(V^A\,_{B;i;j})=[V^A\,_{B;i;j;k}+(S_{Cl}\,^{Am}.V^C\,_{B;i;j}-S_{B \overline{%
l}}\,^{D\underline{m}}.V^A\,_{D;i;j}).T_{km}^l+ \\ 
+V^A\,_{B;l;j}.T_{ki}^l+V^A\,_{B;i;l}.T_{kj}^l].\xi ^k+ \\ 
+(S_{Cl}\,^{Ak}.V^C\,_{B;i;j}-S_{B\overline{l}}\,^{D\underline{k}%
}.V^A\,_{D;i;j}).\xi ^l\,_{;k}+V^A\,_{B;l;j}.\xi ^l\,_{;i}+V^A\,_{B;i;l}.\xi
^l\,_{;j}\text{ .}
\end{array}
\end{equation}

3. Lie derivatives of the components $g_{ij}$ of the metric tensor field $g$%
\begin{equation}  \label{a.4}
\pounds _\xi g_{ij}=(g_{ij;k}+g_{lj}.T_{k\underline{i}}^{\overline{l}%
}+g_{il}.T_{k\underline{j}}^{\overline{l}}).\xi ^k+g_{lj}.\xi ^{\overline{l}%
}\,_{;\underline{i}}+g_{il}.\xi ^{\overline{l}}\,_{;\underline{j}}\text{ .}
\end{equation}
\begin{equation}  \label{a.5}
-S_{B\overline{i}}\,^{D\underline{j}}.g_D=g_k^{\underline{j}}.g_{\overline{i}%
l}+g_l^{\underline{j}}.g_{\overline{i}k}\text{ , \thinspace \thinspace
\thinspace \thinspace }B=kl\text{ ,\thinspace \thinspace \thinspace
\thinspace }D=mn\text{ .}
\end{equation}

\subsection{Energy-momentum tensors for the non-metric tensor fields $%
V^A\,_B $}

1. Energy-momentum tensor of Belinfante $_{sv}T_i$ $^j$ for $V^A$ $_B$%
\begin{equation}  \label{B.1}
_{vs}T_i\text{ }^j=\,_v\mathit{T}_i\text{ }^j-g_i^j.L\text{ ,}
\end{equation}
\begin{equation}  \label{B.2}
_v\mathit{T}_i\text{ }^j=\,_v\mathit{T}_i\text{ }^{jk}\text{ }_k=g_{im}._v%
\mathit{T}^{\overline{m}jk}\text{ }_k\text{ ,\thinspace \thinspace
\thinspace \thinspace \thinspace \thinspace \thinspace \thinspace \thinspace
\thinspace }_v\mathit{T}_i\text{ }^{jk}\text{ }_k=g_k^l._v\mathit{T}_i\text{ 
}^{jk}\text{ }_l\text{ ,}
\end{equation}
\begin{equation}  \label{B.3}
_v\mathit{T}_i\text{ }^{jk}\text{ }_l=g_{im}.(_{vs}\overline{V}_r\text{ }%
^{kj}\text{ }_l.g^{\overline{r}\overline{m}}+\,_{vs}\overline{V}_r\text{ }^{k%
\overline{m}}\text{ }_l.g^{\overline{r}j}-\,_{vs}\overline{V}_r\text{ }^{j%
\overline{m}}\text{ }_l.g^{\overline{r}k})=g_{im}._v\mathit{T}^{\overline{m}%
jk}\,_l\text{ ,}
\end{equation}
\begin{equation}  \label{B.5}
_v\overline{V}_r\text{ }^{kj}\text{ }_l=\,_{vs}\overline{V}_r\text{ }^{kj}%
\text{ }_l+\,_{va}\overline{V}_r\text{ }^{kj}\text{ }_l\text{ ,}
\end{equation}
\begin{equation}  \label{B.6}
_{va}\overline{V}_r\text{ }^{kj}\text{ }_l=\frac 12(_v\overline{V}_r\text{ }%
^{kj}\text{ }_l-\,_v\overline{V}_r\text{ }^{jk}\text{ }_l)\text{ , }_{vs}%
\overline{V}_r\text{ }^{kj}\text{ }_l=\,_{vs}\overline{V}_r\text{ }^{jk}%
\text{ }_l=\frac 12(_v\overline{V}_r\text{ }^{kj}\text{ }_l+\,_v\overline{V}%
_r\text{ }^{jk}\text{ }_l)\text{ ,}
\end{equation}
\begin{equation}  \label{B.7}
_v\overline{V}_r\text{ }^{kj}\text{ }_l=\,_v\overline{Q}_{\overline{r}}\text{
}^{k\underline{j}}\text{ }_l-\,_v\overline{P}_r\text{ }^{kj}\text{ }_l\text{
,}
\end{equation}
\[
_v\overline{P}_r\text{ }^{kj}\text{ }_l=S_{Cr}\text{ }^{Aj}.[\frac{\partial L%
}{\partial V^A\text{ }_{B;k}}.V^C\text{ }_B+(\frac{\partial L}{\partial V^A%
\text{ }_{B;k;m}}+\frac{\partial L}{\partial V^A\text{ }_{B;m;k}}).V^C\text{ 
}_{B;m}- 
\]
\begin{equation}  \label{B.8}
-(\frac{\partial L}{\partial V^A\text{ }_{B;k;m}}.V^C\text{ }_B)_{;m}]_{;l}+(%
\frac{\partial L}{\partial V^A\text{ }_{B;j;k}}.V^A\text{ }_{B;r})_{;l}\text{
,}
\end{equation}
\[
_v\overline{Q}_{\overline{r}}\text{ }^{k\underline{j}}\text{ }_l=S_{B%
\overline{r}}\text{ }^{D\underline{j}}.[\frac{\partial L}{\partial V^A\text{ 
}_{B;k}}.V^A\text{ }_D+(\frac{\partial L}{\partial V^A\text{ }_{B;k;m}}+%
\frac{\partial L}{\partial V^A\text{ }_{B;m;k}}).V^A\text{ }_{D;m}- 
\]
\begin{equation}  \label{B.9}
-\,(\frac{\partial L}{\partial V^A\text{ }_{D;k;m}}.V^A\text{ }_D)_{;m}]_{;l}%
\text{ ,}
\end{equation}
\begin{equation}  \label{B.10}
S_{Cm}\text{ }^{Aj}=-%
\sum_{k=1}^lg_{j_k}^j.g_m^{i_k}.g_{j_1}^{i_1}....g_{j_{k-1}}^{i_{k-1}}.g_{j_{k+1}}^{i_{k+1}}...g_{j_l}^{i_l}%
\text{ ,}
\end{equation}
\begin{equation}  \label{B.11}
V^A\text{ }_{B;i}=e_iV^A\text{ }_B+\Gamma _{Ci}^A.V^C\text{ }_B+P_{Bi}^D.V^A%
\text{ }_D=V^A\text{ }_B,i+\Gamma _{Ci}^A.V^C\text{ }_B+P_{Bi}^D.V^A\text{ }%
_D\text{ ,}
\end{equation}
\begin{equation}  \label{B.12}
\Gamma _{Ci}^A=-S_{Cm}\text{ }^{Aj}.\Gamma _{ji}^m\text{ ,\thinspace
\thinspace \thinspace \thinspace \thinspace }P_{Ci}^A=-S_{Cm}\text{ }%
^{Aj}.P_{ji}^m\text{ . }
\end{equation}

2. Generalized canonical energy-momentum tensor $_v\overline{\theta }_i$ $^j$
for $V^A$ $_B$. $_v\overline{t}_i$ $^j$ is the canonical energy-momentum
tensor for $V^A\,_B$. 
\begin{equation}  \label{B.13}
_v\overline{\theta }_i\text{ }^j=\,_v\overline{t}_i\text{ }^j-\,_vK_i\text{ }%
^j-\,_v\overline{W}_i\text{ }^{jk}\text{ }_k\text{ ,}
\end{equation}
\begin{equation}  \label{B.14}
_v\overline{t}_i\text{ }^j=[\frac{\partial L}{\partial V^A\text{ }_{B;j}}-(%
\frac{\partial L}{\partial V^A\text{ }_{B;k;j}}+\frac{\partial L}{\partial
V^A\text{ }_{B;j;k}})_{;k}].V^A\text{ }_{B;i}+(\frac{\partial L}{\partial V^A%
\text{ }_{B;k;j}}.V^A\text{ }_{B;i})_{;k}-g_i^j.L\text{\thinspace ,}
\end{equation}
\[
_vK_i\text{ }^j=S_{Cm}\text{ }^{An}.V^C\text{ }_B.\frac{\partial L}{\partial
V^A\text{ }_{B;k;j}}.R^m\text{ }_{nik}+S_{Bm}\text{ }^{Dn}.V^A\text{ }_D.%
\frac{\partial L}{\partial V^A\text{ }_{B;k;j}}.P^m\text{ }_{nik}+ 
\]
\begin{equation}  \label{B.15}
+\,\frac{\partial L}{\partial V^A\text{ }_{B;k;j}}.V^A\text{ }_{B;l}.T_{ik}^l%
\text{ ,}
\end{equation}
\begin{equation}  \label{B.16}
\begin{array}{c}
T_{ij}^k=-T_{ji}^k=\Gamma _{ji}^k-\Gamma _{ij}^k-C_{ij} \text{ }^k\,\,\,%
\text{(in a non-co-ordinate basis),} \\ 
T_{ij}^k=\Gamma _{ji}^k-\Gamma _{ij}^k\,\,\,\,\,\,\,\text{(in a co-ordinate
basis),}
\end{array}
\end{equation}
\begin{equation}  \label{B.17}
_v\overline{W}_i\text{ }^{jk}\text{ }_k=g_k^l._v\overline{W}_i\text{ }^{jk}%
\text{ }_l=g_k^l.g_{im}._v\overline{W}^{\overline{m}jk}\text{ }_l\text{ ,}
\end{equation}
\begin{equation}  \label{B.18}
_v\overline{W}^{\overline{m}jk}\text{ }_l=\,_{vs}\overline{V}_n\text{ }^{j%
\overline{m}}\text{ }_l.g^{\overline{n}k}-\,_{vs}\overline{V}_n\text{ }^{k%
\overline{m}}\text{ }_l.g^{\overline{n}j}-\,_{va}\overline{V}_n\text{ }^{jk}%
\text{ }_l.g^{\overline{n}\overline{m}}=-\,_v\overline{W}^{\overline{m}kj}%
\text{ }_l\text{ .}
\end{equation}

3. Variational energy-momentum tensor of Euler-Lagrange $_v\overline{Q}_i$ $%
^j$ for $V^A$ $_B$%
\begin{equation}  \label{B.19}
_v\overline{Q}_i\text{ }^j=(S_{B\overline{i}}\text{ }^{D\underline{j}}.V^A%
\text{ }_D-S_{Ci}\text{ }^{Aj}.V^C\text{ }_B).\frac{\delta _vL}{\delta V^A%
\text{ }_B}\text{ .}
\end{equation}
\begin{equation}  \label{B.20}
_v\overline{F}_i=\frac{\delta _vL}{\delta V^A\text{ }_B}.V^A\text{ }%
_{B;i}+\,_vW_i\text{ ,\thinspace \thinspace \thinspace \thinspace \thinspace
\thinspace \thinspace \thinspace }_vW_i=\,_vS_i-\,_vS_k\text{ }^j.T_{ij}^k%
\text{ , }
\end{equation}
\begin{equation}  \label{B.20a}
_vW_i=\,_vS_i-\,_vS_k\,^j.T_{ij}^k+g_{i;j}^j.L\text{ , \thinspace }
\end{equation}
\begin{equation}  \label{B.20b}
\begin{array}{c}
_vS_i=\,_v \overline{W}_i\,^{jk}\,_{k;j}+\,_v\overline{V}_l\,^{kj}.R^l%
\,_{jik}-\,_v\overline{Q}_{\overline{l}}\,^{k\underline{j}%
}.(g_{j;k;i}^l-g_{j;i;k}^l+g_{j;m}^l.T_{ik}^m)= \\ 
=\,_v \overline{W}_i\,^{jk}\,_{k;j}-\,_v\overline{W}_i\,^{jk}\,_{;k;j}- \\ 
-\frac 12[_v \overline{W}_i\,^{jk}\,_{;l}.T_{jk}^l+\,_v\overline{W}%
_l\,^{jk}.(T_{<ij}\,^l\,_{;k>}+T_{<ij}\,^m.T_{mk>\,}{}^l)+ \\ 
+\,_v \overline{W}_i\,^{jk}.(T_{jk}\,^l\,_{;l}+T_{jk}-T_{kj}+T_l.T_{jk}%
\,^l-R^m\,_{mjk})+ \\ 
+\,_v \overline{W}_l\,^{jk}.(g_{i;j;k}^l-g_{i;k;j}^l+g_{i;m}^l.T_{kj}\,^m)]+
\\ 
+\,_v \mathit{T}^{\overline{m}jk}.g_{ml}.R^l\,_{jik}-\,_v\overline{Q}_{%
\overline{l}}\,^{k\underline{j}}(g_{j;k;i}^l-g_{j;i;k}^l+g_{j;m}^l.T_{ik}%
\,^m)\text{ ,} \\ 
T_i=g_l^k.T_{ik}\,^l=T_{il}\,^l\text{ ,\thinspace \thinspace \thinspace
\thinspace \thinspace \thinspace }T_{jk}=g_m^l.T_{lj}\,^m\,_{;k}\text{ ,}
\end{array}
\end{equation}
\begin{equation}  \label{B.20c}
_v\overline{V}_i\,^{kj}=\,_v\overline{Q}_{\overline{i}}\,^{k\underline{j}%
}-\,_v\overline{P}_i\,^{kj}\text{ ,}
\end{equation}
\[
_v\overline{Q}_{\overline{r}}\,^{k\underline{j}}=S_{B\overline{r}}\text{ }^{D%
\underline{j}}.[\frac{\partial L}{\partial V^A\text{ }_{B;k}}.V^A\text{ }_D+(%
\frac{\partial L}{\partial V^A\text{ }_{B;k;m}}+\frac{\partial L}{\partial
V^A\text{ }_{B;m;k}}).V^A\text{ }_{D;m}- 
\]
\begin{equation}  \label{B.20d}
\,-(\frac{\partial L}{\partial V^A\text{ }_{D;k;m}}.V^A\text{ }_D)_{;m}]%
\text{ ,}
\end{equation}
\[
_v\overline{P}_r\text{ }^{kj}=S_{Cr}\text{ }^{Aj}.[\frac{\partial L}{%
\partial V^A\text{ }_{B;k}}.V^C\text{ }_B+(\frac{\partial L}{\partial V^A%
\text{ }_{B;k;m}}+\frac{\partial L}{\partial V^A\text{ }_{B;m;k}}).V^C\text{ 
}_{B;m}- 
\]
\begin{equation}  \label{B.20e}
-\,(\frac{\partial L}{\partial V^A\text{ }_{B;k;m}}.V^C\text{ }_B)_{;m}]+%
\frac{\partial L}{\partial V^A\text{ }_{B;j;k}}.V^A\text{ }_{B;r}\text{ ,}
\end{equation}
\begin{equation}  \label{B.20f}
\begin{array}{c}
_v \overline{V}_l\,^{kj}=\,_{vs}\overline{V}_l\,^{kj}+\,_{va}\overline{V}%
_l\,^{kj}\text{ ,\thinspace }\, \\ 
_{vs}\overline{V}_l\,^{kj}=\frac 12(\text{\thinspace }_v\overline{V}%
_l\,^{kj}+_v\overline{V}_l\,^{jk})\text{ ,\thinspace \ \ \thinspace }_{va}%
\overline{V}_l\,^{kj}=\frac 12(_v\overline{V}_l\,^{kj}-_v\overline{V}%
_l\,^{jk})\,\text{ ,\thinspace }
\end{array}
\end{equation}
\begin{equation}  \label{B.20g}
_v\overline{V}_i\,^{kj}=\,_v\overline{W}_i\,^{jk}+\,_v\mathit{T}_i\,^{jk}%
\text{ , \thinspace }_v\overline{W}_i\,^{jk}=g_{il}._v\overline{W}^{%
\overline{l}jk}=-\,_v\overline{W}_i\,^{kj}\text{ ,}\,\,_v\mathit{T}%
_i\,^{jk}=g_{il}._v\mathit{T}^{\overline{l}jk}\text{ ,}
\end{equation}
\begin{equation}  \label{B.20h}
_v\overline{W}^{\overline{l}jk}=\,_{vs}\overline{V}_m\,^{j\overline{l}}.g^{k%
\overline{m}}-\,_{vs}\overline{V}_m\,^{\overline{l}k}.g^{j\overline{m}%
}+\,_{va}\overline{V}_m\,^{kj}.g^{\overline{l}\overline{m}}\text{ ,}
\end{equation}
\begin{equation}  \label{B.20i}
_v\mathit{T}^{\overline{l}jk}=\,_{vs}\overline{V}_m\,^{jk}.g^{\overline{l}%
\overline{m}}+\,_{vs}\overline{V}_m\,^{\overline{l}k}.g^{j\overline{m}%
}-\,_{vs}\overline{V}_m\,^{j\overline{l}}.g^{k\overline{m}}\text{ ,}
\end{equation}
\[
_vS_k\,^j=\,_v\overline{t}_k\,^j-\,_vK_k\,^j+g_k^j.L\text{ .} 
\]

\subsection{Energy-momentum tensors for the metric tensor field $g_{ij}$ 
\newline%
}

1. Symmetric energy-momentum tensor of Belinfante $_{sg}T_i$ $^j$ for $%
g_{ij} $%
\begin{equation}  \label{B.21}
_{gs}T_i\text{ }^j=\,_g\mathit{T}_i\text{ }^j-g_i^j.L\text{ ,}
\end{equation}
\begin{equation}  \label{B.22}
_g\mathit{T}_i\text{ }^j=\,_g\mathit{T}_i\text{ }^{jk}\text{ }_k=g_{il}._g%
\mathit{T}^{\overline{l}jk}\text{ }_k\text{ ,}
\end{equation}
\begin{equation}  \label{B.23}
_g\mathit{T}_i\text{ }^{jk}\text{ }_k=g_{il}.(_{gs}\overline{V}_m\text{ }%
^{kj}\text{ }_k.g^{\overline{m}\overline{l}}+\,_{gs}\overline{V}_m\text{ }^{k%
\overline{l}}\text{ }_k.g^{\overline{m}j}-\,_{gs}\overline{V}_m\text{ }^{j%
\overline{l}}\text{ }_k.g^{\overline{m}k})\text{ ,}
\end{equation}
\begin{equation}  \label{B.24}
_{gs}\overline{V}_i\text{ }^{jk}\text{ }_k=g_k^l._{gs}\overline{V}_i\text{ }%
^{jk}\text{ }_l\text{ ,}
\end{equation}
\begin{equation}  \label{B.25}
_{gs}\overline{V}_i\text{ }^{jk}\text{ }_l=\frac 12(_g\overline{V}_i\text{ }%
^{jk}\text{ }_l+\,_g\overline{V}_i\text{ }^{kj}\text{ }_l)\text{ ,
\thinspace \thinspace }_{ga}\overline{V}_i\text{ }^{jk}\text{ }_l=\frac 12(_g%
\overline{V}_i\text{ }^{jk}\text{ }_l-\,_g\overline{V}_i\text{ }^{kj}\text{ }%
_l)\text{ ,}
\end{equation}
\begin{equation}  \label{B.26}
_g\overline{V}_i\text{ }^{jk}\text{ }_l=\,_g\overline{Q}_{\overline{i}}\text{
}^{j\underline{k}}\text{ }_l-\,_g\overline{P}_i\text{ }^{jk}\text{ }_l\text{%
\thinspace ,}
\end{equation}
\begin{equation}  \label{B.27}
_g\overline{Q}_{\overline{i}}\text{ }^{k\underline{j}}\text{ }_l=S_{B%
\overline{i}}\,^{D\underline{j}}.[\frac{\partial L}{\partial g_{B;k}}.g_D+(%
\frac{\partial L}{\partial g_{B;k;m}}+\frac{\partial L}{\partial g_{B;m;k}}%
).g_{D;m}-(\frac{\partial L}{\partial g_{B;k;m}}.g_D)_{;m}]_{;l}\text{ ,}
\end{equation}
\begin{equation}  \label{B.28}
_g\overline{P}_i\text{ }^{jk}\text{ }_l=(\frac{\partial L}{\partial
g_{mn;k;j}}.g_{mn;i})_{;l}=\,_g\overline{P}_i\text{ }^{jk}\text{ }_{;l}\text{
,\thinspace \thinspace \thinspace \thinspace \thinspace \thinspace
\thinspace \thinspace \thinspace }_g\overline{P}_i\text{ }^{jk}=\frac{%
\partial L}{\partial g_{mn;k;j}}.g_{mn;i}\text{\thinspace \thinspace
\thinspace }
\end{equation}
\begin{equation}  \label{B.29a}
_g\overline{Q}_{\overline{i}}\text{ }^{k\underline{j}}=S_{B\overline{i}}\,^{D%
\underline{j}}.[\frac{\partial L}{\partial g_{B;k}}.g_D+(\frac{\partial L}{%
\partial g_{B;k;m}}+\frac{\partial L}{\partial g_{B;m;k}}).g_{D;m}-(\frac{%
\partial L}{\partial g_{B;k;m}}.g_D)_{;m}]\text{ ,}
\end{equation}
\[
_g\overline{V}_i\text{ }^{jk}=\,_g\overline{Q}_{\overline{i}}\text{ }^{j%
\underline{k}}-\,_g\overline{P}_i\text{ }^{jk}\text{\thinspace ,\thinspace
\thinspace \thinspace \thinspace \thinspace }_g\overline{V}_i\text{ }%
^{jk}=\,_{gs}\overline{V}_i\text{ }^{jk}+\,_{ga}\overline{V}_i\text{ }^{jk}%
\text{\thinspace .\thinspace } 
\]

2. Generalized canonical energy-momentum tensor $_g\overline{\theta }_i$ $^j$
for $g_{ij}$. $_g\overline{t}_i$ $^j$ is the canonical energy-momentum
tensor for $g_{ij}$%
\begin{equation}  \label{B.29}
_g\overline{\theta }_i\text{ }^j=\,_g\overline{t}_i\text{ }^j-\,_gK_i\text{ }%
^j-\,_g\overline{W}_i\text{ }^{jk}\text{ }_k\text{ ,}
\end{equation}
\begin{equation}  \label{B.30}
_g\overline{t}_i\text{ }^j=[\frac{\partial L}{\partial g_{kl;j}}-(\frac{%
\partial L}{\partial g_{kl;j;m}}+\frac{\partial L}{\partial g_{kl;m;j}}%
)_{;m}].g_{kl;i}+(\frac{\partial L}{\partial g_{kl;m;j}}%
.g_{kl;i})_{;m}-g_i^j.L\text{ ,}
\end{equation}
\begin{equation}  \label{B.31}
_gK_i\text{ }^j=2.\frac{\partial L}{\partial g_{kl;m;j}}.g_{nl}.R^n\text{ }%
_{kim}+\frac{\partial L}{\partial g_{kl;m;j}}.g_{kl;n}.T_{im}^n\text{ ,}
\end{equation}
\begin{equation}  \label{B.32}
_g\overline{W}_i\text{ }^{jk}\text{ }_k=g_{il}.(_{gs}\overline{V}_m\text{ }%
^{j\overline{l}}\text{ }_k.g^{\overline{m}k}-\,_{gs}\overline{V}_m\text{ }^{k%
\overline{l}}\text{ }_k.g^{\overline{m}j}-\,_{ga}\overline{V}_m\text{ }^{jk}%
\text{ }_k.g^{\overline{m}\overline{l}})\text{ ,}
\end{equation}

3. Energy-momentum of Euler-Lagrange $_g\overline{Q}_i$ $^j$ for $g_{ij}$%
\begin{equation}  \label{B.33}
_g\overline{Q}_i\text{ }^j=-2.\frac{\delta _gL}{\delta g_{\underline{j}k}}%
.g_{\overline{i}k}\text{ ,}
\end{equation}
\begin{equation}  \label{B.34}
_g\overline{F}_i=\frac{\delta _gL}{\delta g_{kl}}.g_{kl;i}+\,_gW_i\text{ ,
\thinspace \thinspace \thinspace \thinspace }_gW_i=\,_gS_i-\,_gS_l\text{ }%
^k.T_{ik}\text{ }^l+(g_i^j.L)_{;j}\text{ ,}
\end{equation}
\begin{equation}  \label{B.36}
_gS_i=\,_g\overline{W}_i\text{ }^{jk}\text{ }_{k;j}+\,_g\overline{V}_l\text{ 
}^{kj}.R^l\text{ }_{jik}-\,_g\overline{Q}_l%
\,^{kj}.(g_{j;k;i}^l-g_{j;i;k}^l+g_{j;l}^l.T_{ik}^l)\text{ ,}
\end{equation}
\begin{equation}  \label{B.37}
_gS_i\text{ }^j=[\frac{\partial L}{\partial g_{kl;j}}-(\frac{\partial L}{%
\partial g_{kl;j;m}})_{;m}].g_{kl;i}+\frac{\partial L}{\partial g_{kl;m;j}}%
.g_{kl;m;i}=\,_g\overline{t}_i\,^j-\,_gK_i\,^j+g_i^j.L\text{ .}
\end{equation}

\begin{center}
\textbf{References}
\end{center}

\begin{enumerate}
\item  {\small Eddington A. S., \textit{Relativit\"{a}tstheorie in
mathematischer Behandlung} (Verlag von Julius Springer, Berlin, 1925).}

\item  {\small Hartley D., \textit{Normal frames for non-Riemannian
connections}. Class. and Quantum Grav. \textbf{12 (}1995) L103-L105.}

\item  {\small Hecht R. D. and Hehl F. W., \textit{A Metric-Affine Framework
for a Gauge Theory of Gravity}},{\small \ in Proc. 9th Italian Conf. on Gen.
Rel. and Grav. Physics. Capri, Italy 1991, eds. Cianci R. et al. (World Sci.
Pub. Co., Singapore, 1991), \textit{\ }pp. 246-291. }

\item  {\small Hehl F. W., McCrea J. D., Mielke E. W. and Ne'eman Y., 
\textit{Metric-affine gauge theory of gravity: field equations, Noether
identities, world spinors, and breaking of dilation invariance}},{\small \
Physics Reports \textbf{258}, 1-2, (1995) 1-171. }

\item  {\small Hutson V. C. L. and Pym J. S. \textit{Applications of
Functional Analysis and Operator Theory}},{\small \ (Academic Press, London,
1980).}

\item  {\small Iliev B. Z., \textit{Special bases for derivations of tensor
algebras. I. Cases in a neighborhood and at a point}},{\small \textit{\ }%
Comm. JINR Dubna E5-92-507\textbf{\ }(1992\textbf{)} 1-17.}

\item  \_\_\_\_\_ ,{\small \ \textit{Special bases for derivations of tensor
algebras}. \textit{II. Case along paths}},{\small \textbf{\ }Comm. JINR%
\textit{\ }Dubna E5-92-508\textbf{\ }(1992\textbf{)} 1-16.}

\item  \_\_\_\_\_ ,{\small \ \textit{Special bases for derivations of tensor
algebras. III. Case along smooth maps with separable points of
selfintersection}},{\small \textbf{\ }Comm. JINR Dubna E5-92-543\textbf{\ }%
(1992\textbf{)} 1-15.}

\item  {\small Ljusternik L. A. and Sobolev V. I., \textit{Concise course of
functional analysis}, (Visshaja shkola, Moscow, 1982), (in Russian). }

\item  {\small Lovelock D. and Rund H., \textit{Tensors, Differential Forms,
and Variational Principles}},{\small \ (John Wiley \& Sons, New York, 1975). 
}

\item  {\small Manoff S., \textit{On the energy-momentum tensors for
theories in (pseudo) Riemannian spaces with torsion}, Comm. JINR Dubna
E2-87-679 (1987) 1-16. }

\item  \_\_\_\_\_ ,{\small \ \textit{On the energy-momentum tensors for
field theories in spaces with affine connection and metric. Generalized
Bianchi Identities and Different Energy-Momentum Tensors}},{\small \ Comm.
JINR Dubna E2-91-77 (1991) 1-16.}

\item  \_\_\_\_\_ ,{\small \ \textit{On the energy-momentum tensors for
field theories in spaces with affine connection and metric. Conditions for
Existence of Energy-Momentum Tensors}},{\small \ Comm. JINR Dubna E2-91-78
(1991) 1-15.}

\item  {\small \_\_\_\_\_ , \textit{Kinematics of Vector Fields}},{\small \
in \textit{Complex Structures and Vector Fields}, eds. Sekigawa K., Dimiev
S. (World Scientific Publ., Singapore, 1995), pp. 61-113.}

\item  {\small \_\_\_\_\_ , \textit{Geodesic and autoparallel equation over
differentiable manifolds}},{\small \ Intern. J.\ Mod. Phys. A. \textbf{11},
21 (1996) 3849-3874.}

\item  {\small Schmutzer E., \textit{Relativistische Physik (Klassische
Theorie)}},{\small \ (B. G. Teubner Verlagsgesellschaft, Leipzig, 1968). }

\item  {\small Schr\"{o}dinger E., \textit{Space-Time Structure,} (Cambridge
at the University Press, Cambridge, 1950\textbf{).} }

\item  {\small von der Heyde P., \textit{The Equivalence Principles in the }$%
U_4$\textit{\ Theory of Gravitation}},{\small \textbf{\ }Lett. Nuovo Cim. 
\textbf{14,} 7 \textbf{(}1975\textbf{) }250-252.\ }

\item  {\small Yano K., \textit{The Theory of Lie Derivatives and Its
Applications}},{\small \ (North-Holland Publ. Co., Amsterdam, 1957). }

\item  {\small Zoritch V. A., \textit{Mathematical analysis}. Vol.1.,
(Nauka, Moscow, 1981), (in Russian).}
\end{enumerate}

{\small \ }\textsc{Bulgarian Academy of Sciences, Institute for Nuclear
Research and Nuclear Energy, Department of Theoretical Physics, Blvd.
Tzarigradsko Chaussee 72, 1784 Sofia - Bulgaria}

\textit{E-mail address}: smanovf@inrne.bas.bg

\end{document}